\documentclass{aa}  
\usepackage{graphicx}
\usepackage{txfonts}
\newcommand{\HI}{\textup{H\,{\mdseries\textsc{i}}}~}
\newcommand{\HII}{\textup{H\,{\mdseries\textsc{ii}}}}

\newcommand{\kms}{{km~s$^{-1}$}}

\newcommand{\HaN}{{H$\alpha$+[N II]}\ }
\newcommand{\Ha}{{H$\alpha$}\ }

\newcommand{\Msun}{{M$_{\odot}$}}

\def\kms{{km~s$^{-1}$}}

\def\arcmin{\hbox{$^\prime$}}
\def\arcsec{\hbox{$^{\prime\prime}$}}

\def\farcmin{\hbox{$.\mkern-4mu^\prime$}}
\def\farcsec{\hbox{$.\!\!^{\prime\prime}$}}

\begin{document}

 \title{Dorado and its member galaxies. III   }

\subtitle{Mapping star formation with FUV imaging from {\tt UVIT}.
\footnote{Data set are only available via anonymous ftp to cdsarc.u-stradbg.fr 
	(130.79.128.5).}  }

   \author{R. Rampazzo
          \inst{1,2}
          \and
   P. Mazzei\inst{2}
                        \and
   {A. Marino}\inst{2}
\and
       {L. Bianchi}\inst{3} 
\and 
       {J. Postma}\inst{4}
\and \\
{R. Ragusa}\inst{5}
\and
       {M. Spavone}\inst{5} 
       \and
             {E. Iodice}\inst{5}
       \and
         {S. Ciroi}\inst{6}
                 \and
         {E.V. Held}\inst{2}
      }

\institute{INAF-Osservatorio Astrofisico di Asiago,Via dell'Osservatorio 8, 36012 Asiago, Italy
  \email{roberto.rampazzo@inaf.it}
   \and
   {INAF-Osservatorio Astronomico di Padova}, {Vicolo dell'Osservatorio 5},  {35122 Padova}, {Italy}
    \and
    {Dept. of Physics \& Astronomy}, {The Johns Hopkins University}, { 3400 N. Charles St., Baltimore}, {MD 21218, USA}
     \and
{University of Calgary}, {2500 University Drive NW, Calgary}, {Alberta, Canada}
    \and
    {INAF-Osservatorio Astronomico di Capodimonte}, {Salita Moiariello 16}, {80131 Napoli}, {Italy}
\and
   {Department of Physics and Astronomy  ``G. Galilei'', University of Padova}, {Vicolo dell'Osservatorio 3}, {35122 Padova}, {Italy}
%
    }

   \date{April 6, 2022 ; May 30, 2022 }


\abstract
{We are investigating the star formation in galaxies of the actively evolving Dorado 
group where, for a large fraction of both early-  and late-type galaxies, 
signatures of interactions and merging events are revealed by optical 
and radio observations.}
{Our previous \HaN\ study, probing $\approx$10 Myrs timescales,  suggested that 
star formation is still ongoing in early-type galaxies. 
In this work,  we use far-UV (FUV) imaging to map recent  star
formation on longer times scales, of the order of 100 Myrs.}
{We used the Ultraviolet telescope {\tt UVIT} on board 
{\tt Astrosat}  to image the galaxies of the Dorado backbone previously observed 
in \HaN, with  the far-UV  filter FUV.CaF2 (1300-1800 \AA).
The sample includes NGC 1536, NGC  1546,
NGC 1549, [CMI2001]4136-01, NGC 1553, IC 2058, PGC 75125,NGC 1566,
NGC~1596 and NGC~1602; for the two latter galaxies, the UVIT data provide the first 
view in far-UV. For the others, previously observed by GALEX, the UVIT data afford 
a $\sim 5 \times$ improvement in spatial resolution.} 
{FUV.CaF2 emission is revealed in all the Dorado galaxies observed, tracing 
young stellar populations in ring structures and showing  tidal distortions. 
The S\'ersic index, derived by fitting the luminosity profiles, is always $n<3$ 
suggesting that the FUV.CaF2 emission originates from a disk also in early-type galaxies.
The star formation rate (SFR) ranges from 
0.004$\pm$0.001~M$_\odot$yr$^{-1}$ of [CMI2001]4136-01 to 
2.455$\pm$0.027~M$_\odot$yr$^{-1}$  of NGC 1566.
Most of the recent star formation is found at the periphery of
the Dorado group where most of late-type galaxies are located.
For these galaxies, the ratio SFR$_{H\alpha}$/SFR$_{FUV.CaF2}$ is  close to 1, 
except for the edge-on IC 2058, similarly to previously reported relations 
for Local Volume samples.  For early-type galaxies, however,  SFR$_{H\alpha}$ is about 15 times higher 
than SFR$_{FUV}$. The Dorado's early-type galaxies  define a separate locus in 
SFR$_{FUV}$,  SFR$_{H\alpha}$ space  with respect to the late-type galaxies,  
which is well represented by the relation 
log (SFR$_{FUV.CaF2})$ = 0.70~$\times$~log (SFR$_{H\alpha})$$-1.26$.} 
{The disk structure of the FUV.CaF2 emitting populations discovered in all
the early-types galaxies implies dissipative processes and wet
merging events. The systematic discrepancy between SFRs
derived from \Ha\ and FUV fluxes suggests  that rejuvenation
episodes in early-type galaxies cannot sustain constant star formation
over $\sim$100 Myrs timescales.}

\keywords{Ultraviolet: galaxies -- Galaxies: elliptical and lenticular, 
	cD -- Galaxies: spiral -- Galaxies: interaction -- Galaxies: evolution }

\maketitle
%

\section{Introduction}

In the  Local Supercluster (LS, $\leq$3500 \kms) 
\citep{deVaucouleurs1953} galaxy groups play an important 
role in the galaxy evolution. \citet{Kourkchi2017} 
suggested that, considering the infall domain of LS,  
only a small fraction of the galaxy mass resides within
clusters. Virgo, Fornax and Antlia 
associations with an infall mass of 7$\times$10$^{14}$ \Msun, 
2$\times$10$^{14}$ \Msun\ and 4$\times$10$^{14}$ \Msun\, respectively, 
contain 60\%, 30\% and 15\% of their mass already collapsed within the cluster
\citep{Kourkchi2017}. 
This evidence suggests that a large fraction of galaxies in the LS
are  still evolving within small galaxy associations, making the 
study of groups critical to understand galaxy evolution.

Before groups fall into clusters, they collapse 
under gravity forming substructures where their galaxy members 
undergo a morphological transformations, as for early-type galaxies (ETGs),
and start to quench.  A number of studies shows that ETGs are more
strongly clustered than late-type galaxies (LTGs) \citep{Davis1976,Dressler1980}.
\citet{Boselli2014} reviewed some of the mechanisms driving galaxy 
evolution as a function of the environment. Such mechanisms act
transforming field galaxies, i.e. LTGs, into cluster-like galaxies, i.e.
ETGs, and drive groups from an {\it active} (star forming) phase, 
typical of the field, \citep[see
e.g.][]{Marino2016,Rampazzo2018} to a more {\it passive} phase, typical 
of the clusters. It is widely admitted that cluster galaxies
tend to have depressed star formation (SF) in comparison to the field
\citep[see e.g.][and references
therein]{Bressan2006,Poggianti2006,Rampazzo2013}. 
At least part of the galaxy transformation happens at densities typical
of the groups, i.e. groups operate a galaxy pre-processing.

We are investigating galaxy groups, from Local Group Analogs 
\citep{Marino2010,Marino2013,Marino2014} to very 
evolved group, as NGC 5486 groups \citep{Marino2016}
using a multi-wavelength approach to understand
their evolution exploring the connection between their galaxy 
population and their activity. This paper continues our study, 
in optical \citep{Cattapan2019} in \Ha\ \citep[][hereafter Ram2020]{Rampazzo2020} 
and FUV \citep[hereafter Ram2021]{Rampazzo2021}, 
of the Dorado group and its sub-structures.\\

\begin{figure}
	{\includegraphics[width=8.9cm]{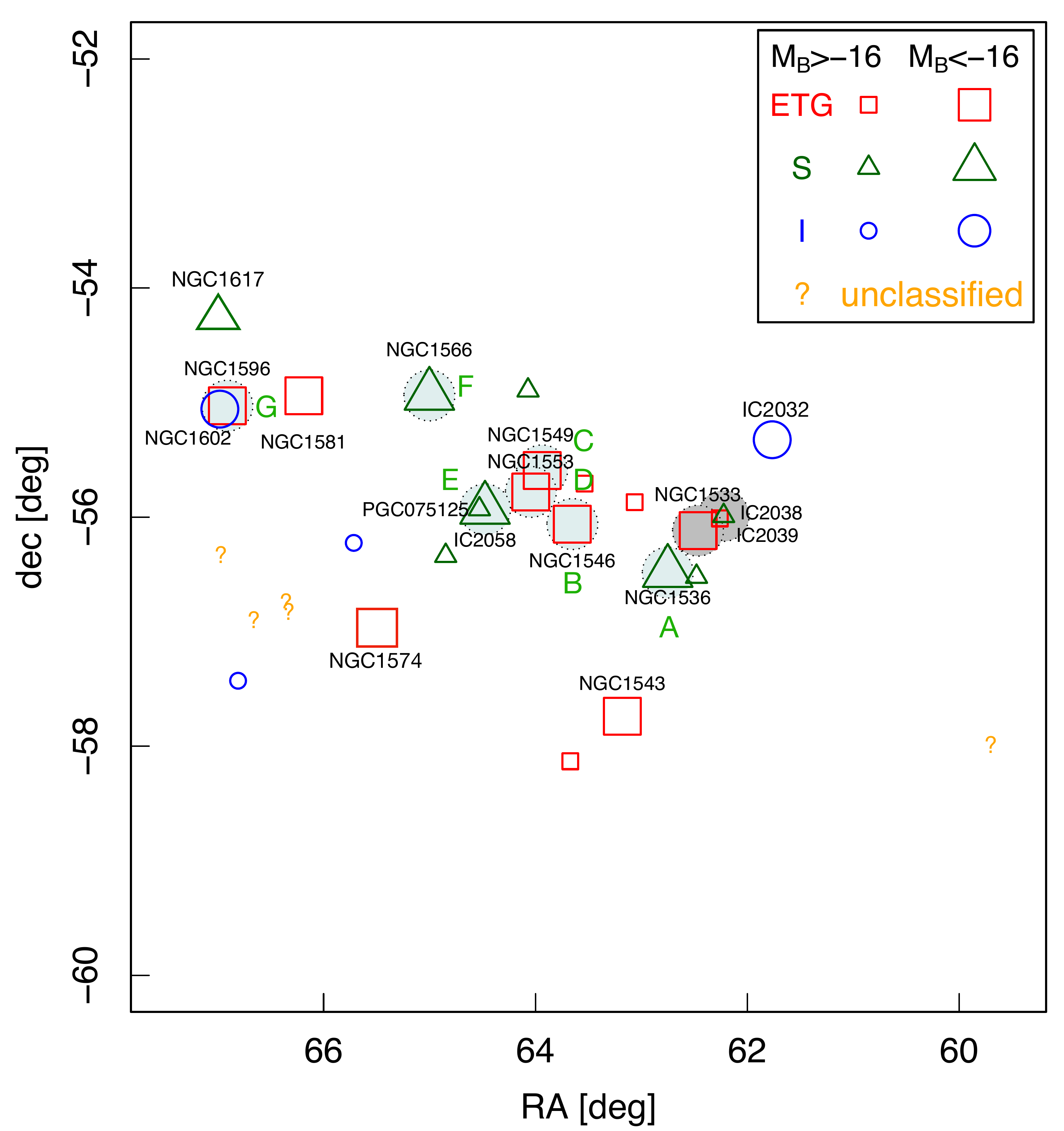}}
	\caption{Projected distribution of Dorado galaxy members according to
\cite{Firth2006} and \citet{Kourkchi2017}. Members are labelled according to their morphology
		and B-Band magnitude provided by {\tt HyperLeda}.  
		{\tt UVIT} FUV.CaF2 fields listed in Table~\ref{UVIT-sources}, are enclosed 
		by a light-green circle having the diameter of the {\tt UVIT} field (28\arcmin). 
		Target galaxies within each field are also labelled.
		The morphological, photometric and star forming properties
		obtained from the {\tt UVIT} FUV.CaF2 data-set about NGC 1533, IC 2038/2039, 
		dark-grey circles in the figure, have been presented and discussed in 
		in Ram2021. Ram2020 
		also observed NGC 1581 and NGC 1543 in \HaN. All galaxies but 
		the NGC~1596/NGC ~1602 have been already observed with {\tt GALEX} 
		at lower resolution and sensitivity \citep[see e.g.][]{Marino2011a,Marino2011b}.}
	\label{Dorado_map}
\end{figure}

Dorado is a nearby  (17.69 Mpc) group in the Southern Hemisphere
(RA=64.3718 deg, Dec=-55.7811 deg). We adopt the member  list
defined by \citet[][and reference therein]{Firth2006} and
\citet{Kourkchi2017} (see Table~A.1 of Ram2020).
Figure~\ref{Dorado_map} shows the projected distribution of Dorado
members, indicating their morphological class and B-band magnitude.
Multi-wavelength data, from  FUV to optical and radio, converge in presenting 
Dorado as a strongly evolving group of galaxies in the LS. 
A large fraction of Dorado members, including ETGs, shows either 
merging or interaction signatures in their
structures (Section~2 in Ram2020). Residual SF
is seen in bright ETGs that populate the red sequence as well as
in  intermediate luminosity members still crossing the green
valley of the UV-optical colour magnitude diagram 
of the group \citep[see Figure 1 in][]{Cattapan2019}. \\

\begin{center}
	\begin{table*}[ht]%
		\centering \tiny
		\caption{Dorado galaxies observed with {\tt UVIT} }%
		\label{UVIT-sources}
		\tabcolsep=0pt%
		\begin{tabular*}{40pc}{@{\extracolsep\fill}lcccccccccc@{\extracolsep\fill}}
			\hline
			\textbf{Field} &\textbf{ID}&  \textbf{RA}& \textbf{Dec}& \textbf{{\tt GALEX} FUV}& \textbf{V$_{hel}$}& \textbf{Morpho.} & \textbf{$M_B$} & \textbf{R$_{25}$}& \textbf{PA} & \textbf{$\epsilon$}\\
			\textbf{centre}&  \textbf{} &  \textbf{(J2000)}     & \textbf{(J2000)} & \textbf{[AB mag]} & \textbf{km s$^{-1}$} & \textbf{Type} &  & arcsec & \textbf{deg} & \\
			\hline
			\hline
			A     &  {\bf NGC 1536}       &  04 11 00.53 &  -56 29 05.6 & 16.03$\pm$0.01$^1$& 1296 & 5.0  & -17.80 & 49.8 & 162.6 & 0.28\\
			\hline
			B &  {\bf NGC 1546}   &  04 14 36.46 & -56 03 39.2    &  17.36$\pm$0.02$^1$&  1238  & -0.4 & -19.24 & 111.5 & 144.8 & 0.32\\
			\hline
			C &  {\bf NGC 1549}      & 04 15 45.13 &-55 35 32.1&   17.06$\pm$0.05$^3$ & 1202 & -4.3 & -20.61 & 153.9 & 146.2 & 0.15 \\
			& [CMI2001]4136-01  & 04 16 15.43   & -55 41 49.2      &       \dots      &  \dots   &\dots      &  \dots & \dots & \dots & \dots \\
			\hline
			D   &   {\bf NGC 1553}   & 04 16 10.50 & -55 46 49.0  & 16.73$\pm$0.05$^3$   & 1201  & -2.3 & -21.02 & 189.3  & 150.4 & 0.31\\
			&      &   &    & 16.32$\pm$0.13$^2$   & 1201  & -2.3 & -21.02\\
			\hline
			E  & {\bf IC 2058}           &   04 17 54.30 &    -55 55 58.0  &   16.15$\pm$0.03$^2$  &   1397  &  6.5 & -17.55 & 95.0 & 17.9 & 0.85\\
			&          &    &    0  &   16.24$\pm$0.02$^4$  &     &   & \\
			
			& {\bf PGC 075125}  & 04 18 07.10 &-55 55 50.0   &   ....  &    1369 &  5.0 & -15.81 & 16.9 & 27.8 & 0.26\\
			\hline
			F &{\bf NGC 1566}    &04 20 00.42  &-54 56 16.1   &  12.13$\pm$0.02$^4$  &    1504 & 4.0 & -20.89 & 217.3 & 44.2 & 0.32\\
			\hline
			G &{\bf NGC 1596}    &  04 27 38.1 & -55 01 40.1 &   \dots    &  1510       &     -2.0 & -19.27 & 116.7 & 19.3 & 0.73\\
			G &{\bf NGC 1602}   &   04 27 54.9 & -55 03 28.1 &   \dots    &  1740           &  9.5 & -17.87 & 55.9 & 88.5 & 0.48   \\       
			\hline
		\end{tabular*}
\tablefoot{Col.1 gives the {\tt UVIT} fields; Col. 2 the galaxy identification. 
Members of Dorado, according to \citet{Kourkchi2017} are indicated in bold face; 
Col.s 3 and 4 give Right Ascension and Declination; Col. 5  the FUV apparent
asymptotic magnitude, not extinction corrected, from:  ($^1$) {\tt GALEX}  
archive, \citet{Bai2015}, ($^3$) Simonian \& Martini (2017) and ($^4$) 
\citet{Gil2007}. Col. 6, Col.7 and Col. 8 lists the heliocentric radial
velocity,  the galaxy morphological type and the absolute B-band
magnitude, respectively by \citet{Kourkchi2017} which adopted for all
the galaxies the  distance of 17.69 Mpc. Col. 9, Col. 10 and 
Col. 11 provide the radius at $\mu_B$=25 mag~arcsec$^{-2}$,
and the corresponding Position Angle and ellipticity 
from {\tt Hyperleda}. The source indicated with
[CMI2001]4136-01 has been already revealed by {\tt GALEX} and  detected in  \HaN\
by Rampazzo (2020), showing that it is a  member of	Dorado.}
	\end{table*}
\end{center}

Dorado has a clumpy structure \citep{Firth2006} 
with respect to evolved groups, e.g. NGC 5486 \citep{Marino2016}. 
\citet{Iovino2002} found that 
the central part of Dorado forms a compact group, namely SGC~0414-5559, 
composed of the brightest galaxies at the peak of the velocity distribution 
(Table A.1 in Ram2020). This structure, 
the core of Dorado, is formed by NGC 1546, NGC 1549, NGC 1553
and IC 2058.Compact groups are the ideal sites to study   
galaxy interactions and mergers given the high space density 
and low velocity dispersions of their galaxies \citep{Mamon1992}. From N-body simulations 
\citet{Diaferio1994} proposed that compact groups continuously form within
loose groups during their collapse and virialization phase. There are several
observational evidences that compact groups are found within looser
structures or rich neighbourhoods \citep[see e.g.][and references
therein]{Ribeiro1998}. \citet{Diaferio1994} simulations suggested that
the life time of the compact groups is about 1 Gyr and most of the simulated member
galaxies are not merger remnants.  More recently, semi-analytic 
simulations (SAMs) by \citet{Diaz2021} suggested a late assembly for 
the compact groups which may host merger remnants. 
This could be the case of SGC~0414-5559 in Dorado, since both
NGC 1549 and NGC 1553 show a wide shell system \citep{Malin1983}.

Some peripheral sub-substructures have been evidenced in Dorado,
as the triplet formed by NGC 1533, IC~2039 and IC~2039 \citep{Cattapan2019},
South-East of the compact group. Pair-like structures as 
NGC 1566/NGC 1581 \citep{Kendall2015,Oh2015} and NGC 1596/NGC 1602
\citep{Bureau2006,Chung2006} are found North-West of the compact group.
Most of these galaxies show signatures of interaction 
as discussed by Ram2020. \\

A further indication that Dorado is at an early stage of evolution 
is the fact that it is still gas-rich when compared to evolved environments.
The basic ingredient for SF, \HI reservoirs, have
been  revealed in several  Dorado members
\citep{Ryan-Weber2003,Kilborn2009,Elagali2019}, with the exclusion of 
NGC 1549 and NGC 1553. There are signatures of gas
transfer between the LTG NGC 1602 and the ETG NGC 1596 \citep{Chung2006}. 
The fuel to ignite SF in the ring of the 
lenticular galaxy NGC 1581 (Ram2020) 
may have been stripped by the interaction \citep{Oh2015,Kendall2015} 
with NGC 1566. 

Ram-pressure stripping signatures of \HI\ have been
detected in NGC 1566 \citep{Elagali2019}. This is particularly
interesting since the galaxy is located at the periphery of the Dorado group.  
 The presence of ram-pressure stripping has long been suggested 
to operate in removing gas from spirals in rich clusters
\citep{Gunn1972} but its efficiency was questioned in less 
rich environments \citep{Abadi1999}. \citet{Kantharia2005} revealed 
signatures in the radio continuum and 21 cm
HI observations of both tidal interaction and ram-pressure stripping on spiral members
of Holmberg~124, a poor group environment. \citet{Roberts2021}
extended the search for jelly-fish galaxies in low density environments
in the radio continuum domain with the LOFAR Two-metre Sky Survey 
(LoTSS) project investigating about 500 SDSS groups (z$<$0.05).
They find that jelly-fish galaxies are most commonly found in clusters, 
with the frequency decreasing towards the lowest-mass groups.
\citet{Vulcani2021} provided a panorama about different 
processes taking place in low-density environments,
including ram-pressure stripping, with VLT/MUSE optical data.
The {\it GALaxy evolution EXplorer}
({\tt GALEX} hereafter) \citep{Martin2005,Morrissey2007} was used to 
identify asymmetries, as tails and jelly-fish tentacles, 
by \citet{Smith2010} in Coma galaxies,
as well as to study star formation in jelly-fish tentacles
in single objects in Virgo \citep[see e.g.][references therein]{Hester2010}. 
More recently, high spatial resolution FUV studies using
Astrosat-{\tt UVIT} investigated  ram-pressure stripping in action 
in cluster jelly-fish galaxies \citep{George2018,Hota2021}.\\

This paper presents our study of the Dorado backbone
based on {\tt Astrosat-UVIT } FUV.CaF2 observations.
We investigate the central compact group SGC~0414-5559 
and the two sub-structures formed by NGC 1566/NGC 1581 and NGC 1596/NGC 1602.
Target galaxies are listed in Table~\ref{UVIT-sources}.
We associated to the FUV.CaF2 images  two deep optical, $g$ and $r$- bands
wide filed images of the Dorado group from  the {\tt ESO-VLT Survey Telescope} 
\citep[{\tt VST} hereafter,][]{Schipani2012}. 
Such images were obtained within the {\tt VEGAS}-survey of nearby ETGs 
\citep{Capaccioli2015,Spavone2017}. The plates used reach
a surface brightness  of  30.5  mag\,arcsec$^{-2}$  
in $g$ and 29.0  mag\,arcsec$^{-2}$  $r$ bands  with a 
resolution of (0\farcsec21/px)  and
a total field of view of 1$^\circ\times1^\circ$. 
The full analysis of the optical data will be presented in a forthcoming 
companion paper (Ragusa et al. 2022, in preparation).
We use {\tt VST} images in order to compare 
the optical R$_{g29}$ and R$_{r28}$ galaxy radii at 
$\mu_g$=29 mag~arcsec$^{-2}$ and $\mu_r$=28 mag~arcsec$^{-2}$,
respectively, with the distribution 
and the extension of the FUV.CaF2  emission.  
	
The plan of the paper is the following.   Section~\ref{Targets} presents
target galaxies and provides the motivation of the {\tt UVIT} study.
In \S~\ref{Observations} we present FUV.CaF2 observations and the 
reduction techniques adopted. The morphological and photometric 
analysis performed and determination of star formation rate (SFR) 
from the FUV.CaF2 integrated galaxy luminosity are
detailed in \S~\ref{Data-Analysis}.  The FUV.CaF2 properties of
individual galaxies are presented and summarized in \S~\ref{Results}. 
We discuss in \S~\ref{Discussion} the FUV.CaF2 versus optical extensions 
of our target galaxies. In Section~\ref{FUV-SFR} we
discuss and compare the SFR from the FUV.CaF2 flux with the values 
obtained from the \Ha\ flux by Ram2020.   
We summarize our results and draw some conclusions 
in \S~\ref{Conclusions}.

\section{The target galaxies and motivation of the FUV study}
\label{Targets}

Ram2020 used  \HaN\ imaging to derive the
SFR of the Dorado backbone galaxies
consisting of 6 LTGs, namely IC~2038, NGC~1536, IC~2058, 
PGC 75125, NGC 1566, NGC 1602 and 8 ETGs, from Es to S0-a 
with a large fraction of S0s, namely IC~2039, NGC~1533, NGC 1543,
NGC 1546, NGC 1549, NGC 1553, NGC 1581 and NGC 1586.  The LTG 
SFR ranges from 0.01$\pm$0.001 
to 2.2$\pm$0.2 M$_\odot$~yr$^{-1}$. These values suggested that LTGs 
in Dorado have a large spread in the SFR if compared with a general LTG sample 
\citep[see e.g.][]{James2004}. The SFR of ETGs ranges from 
0.008$\pm$0.003 to 1.01$\pm$0.10 M$_\odot$~yr$^{-1}$. The comparison  
with the ETGs sample of \citet{Gavazzi2018}, suggested
that the SF in Dorado ETGs are not yet shut down.
Ram2020 proposed that mechanisms such as gas
stripping and gas accretion, through galaxy-galaxy interaction, are
relevant in this evolutionary phase of Dorado.\\

The measure of the SFR from \HaN\ imaging for ETGs is challenging 
due to the low \Ha\ emission fluxes, that need both a careful continuum
subtraction and [NII] line correction
\citep[see e.g.][]{Gavazzi2018,Karachentsev2013}. 
This motivated us to complement the Ram2020 study 
probing the SFR using FUV.CaF2 (1300-1800 \AA) observations 
obtained with {\tt ASTROSAT-UVIT} of the same
Dorado members. {\tt UVIT} observations 
have been used in Ram2021  to study the South West
part of the group, which includes the sub-structure composed of 
NGC 1533,  IC~2038 and IC 2039 (see Figure~\ref{Dorado_map}). 

This paper completes the {\tt UVIT} FUV.CaF2 study of the Dorado backbone.
Our targets map the Dorado central region,
the SGC~0414-5559 compact group together with NGC 1536,
and its North-East extension (Figure~\ref{Dorado_map}) dominated by  
the pairs NGC 1566/NGC 1581 and NGC 1596/NGC 1602.  
Table~\ref{UVIT-sources} provides some relevant properties of the
galaxies here considered. All targets, except for NGC~1596 and NGC~1602,  
had been previously imaged in UV (FUV and  NUV) with {\tt GALEX},   
in a broad  passband similar to FUV.CaF2 as the {\tt UVIT} filter used here, but at about 
5 times lower resolution. Therefore, the new {\tt UVIT} data are a significant 
improvement in resolution  with respect to {\tt GALEX} UV 
studies, as they probe the FUV morphology on about 0.14 kpc scale, the 
typical size of SF complexes, and they are first FUV view of 
these two galaxies.\\ 

\section{Observations and data reduction}
\label{Observations}

{\tt Astrosat} is a X-ray - UV observatory  launched by
the Indian Space Research Organization on September 28, 2015. 
The Ultra-Violet Imaging Telescope facility {\tt UVIT} (Tandon et al. 2017)
is composed of two Ritchey-Chretien telescopes with  37.5 cm aperture,  a circular
field of view of 28\arcmin\ diameter, originally observing simultaneously one in FUV
(1300-1800 \AA) and the other  both in Near UV (NUV) (2000-3000 \AA) and optical
band, VIS (3200-5500 \AA), by means of a beam-splitter directing NUV and
VIS to individual cameras.

\begin{center}
	\begin{table*}
		\centering
		\caption{{\tt UVIT} observations \label{UVIT-observations}}%
		\tabcolsep=0pt%
		\begin{tabular*}{30pc}{@{\extracolsep\fill}lccccc@{\extracolsep\fill}}
			\hline
			\textbf{Field} &  \textbf{Obs ID} &  \textbf{Observing date} & \textbf{Exp. Time}  & \textbf{Target}  \\
			\textbf{ID }          &  \textbf{}    &  \textbf{}     & \textbf{[s]}            &   \textbf{ID }  \\
			\hline
			A  & A07 3238  & October 16, 2019     &  6627.635  & NGC 1536 \\
			B  & A07 3240  & October 16, 2019     &  6669.294 & NGC 1546  \\
			C  & A05 2460  & October 26, 2018     & 3311.534  & NGC 1549  \\
			D  & A05 2460  & October 27, 2018     & 3354.349  & NGC 1553  \\
			E  & A05 2458  & October 26, 2018     & 3189.382  & IC 2058   \\
			F  & G06\_87    & December 26, 2016 &  2940.386 & NGC 1566  \\
			G  & A07 3582  &  March 21, 2020       &  2819.043  & NGC 1596\\
			\hline
		\end{tabular*}
		\tablefoot{Field identification in {\bf Col. 1} refers to the programs A05$\_$02 and A07$\_$010 (PI. R. Rampazzo)
listed in Col. 2.  In Col. 3 and Col. 4  we report the observing date and the 
total effective exposure time.  Col. 5 gives the central target. 
		}
	\end{table*}
\end{center}

The NUV detector is currently not   working. Therefore, observations
have been performed with the FUV channel only. We used  the full field
of view, in photon counting mode, with the Filter F148W CaF2
($\lambda_{mean}$=1481, $\Delta \lambda$=500\AA).  Photons are counted
on a planar CMOS array at approximately 28~Hz and stacked to reconstruct 
the image \citep[see for details][] {Postma2011,Kumar2012,Postma2017, Tandon2017}
with the astrometric world coordinate solution 
solved automatically by a trigonometric algorithm \citep{Postma2020}. 

Table~\ref{UVIT-observations} compiles the relevant information on
the  {\tt Astrosat-UVIT} observations obtained in two programs
A05\_002 and A07\_010 (PI R. Rampazzo).
The first covers the three members of the compact group SGC~0414-5559,
i.e. NGC 1549 (Field C), NGC 1553 (Field D) and IC 2058 (Field E). 
The A07\_010 program observed  the fourth member of the compact group  
NGC 1546 (Field B) and NGC 1536 (Field A), located South of the NGC 1533 
sub-structure (see Ram2020). NGC 1536 has a recession
velocity similar to that of the compact group. Observations
of NGC 1566 (Field F) have been obtained from the {\tt ASTROSAT} 
archive (program G06\_087 PI S. Stalin). We used only the FUV.CaF2 
observation for consistency with our programs.

\section{Data Analysis}
\label{Data-Analysis}

Images reconstructed from the photon-counting data with 0\farcsec416   
subsampling have been rebinned for our analysis, to enhance S/N, to
1\farcsec664 px$^{-1}$. The  nominal zero point magnitude of the 
FUV.CaF2 filter is 18.08\,mag \citep{Tandon2017}.

Surface photometry has been performed using {\tt ELLIPSE}  
fitting routine in the STSDAS package of
IRAF \citep{Jedrzejewski1987}, increasing the size of 
the apertures logarithmically. Foreground and background objects 
have been masked. 
The present {\tt  UVIT} FUV.CaF2 observations allow to measure 
surface brightness profiles down to $\mu_{FUV}\sim29-30$
mag~arcsec$^{-2}$ with an error of $\pm$0.3-0.5 mag~arcsec$^{-2}$.

We applied the procedure outlined by \citet{Ebeling2006}, 
{\tt ASMOOTH}, to enhance the signal-to-noise (S/N) 
ratio  in the galaxy outskirts, in order to 
bring out faint structures in the UV image.
The only parameter required by the procedure is the 
minimum S/N, $\tau_{min}$. 
The algorithm increases the smoothing scale until the 
S/N within the kernel reaches the $\tau_{min}$ input value. 
{\tt ASMOOTH} suppresses very efficiently the noise while
the signal, locally significant at the selected S/N level, is preserved
on all scales. 

\begin{figure}
	{\includegraphics[width=8.9cm]{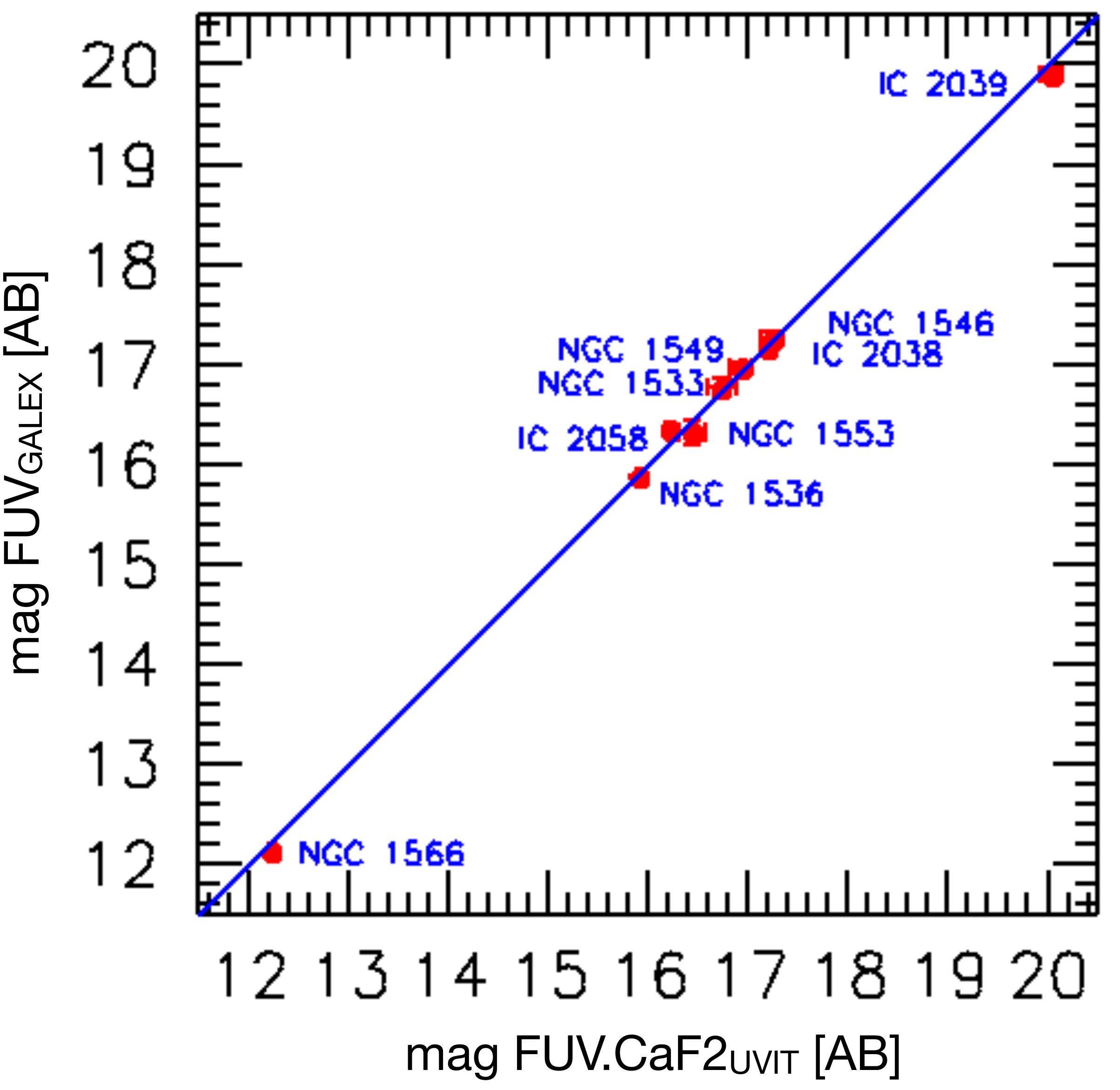}}
	\caption{Comparison with FUV {\tt GALEX} magnitudes 
reported in Table~\ref{UVIT-sources}.
The comparison includes galaxies in Ram2021, namely
NGC 1533, IC~2038 and IC 2039, whose magnitudes are corrected
for foreground Galactic extinction as described in their Table 3.}
	\label{comparison_mag}
\end{figure}
%
%
%
%
\begin{figure}
	{\includegraphics[width=8.9cm]{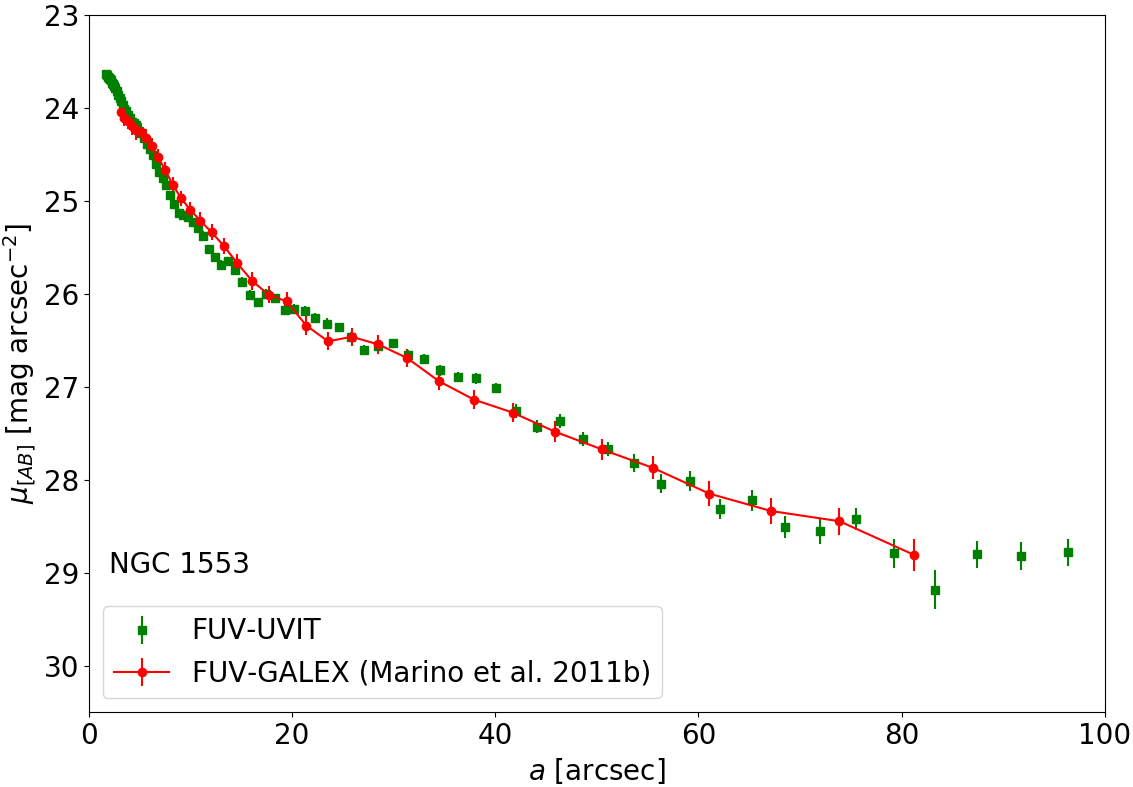}}
	\caption{ {\tt UVIT} FUV.CaF2 luminosity profile of NGC 1553 (green squares) 
compared with that of \citet{Marino2011b} from FUV-{\tt GALEX} observations. 
		The {\tt UVIT} image of NGC 1553 was binned 4$\times$4 pxs.
		\label{NGC1553-comparison}}
\end{figure}

\begin{figure*}
\center
	{\includegraphics[width=13cm]{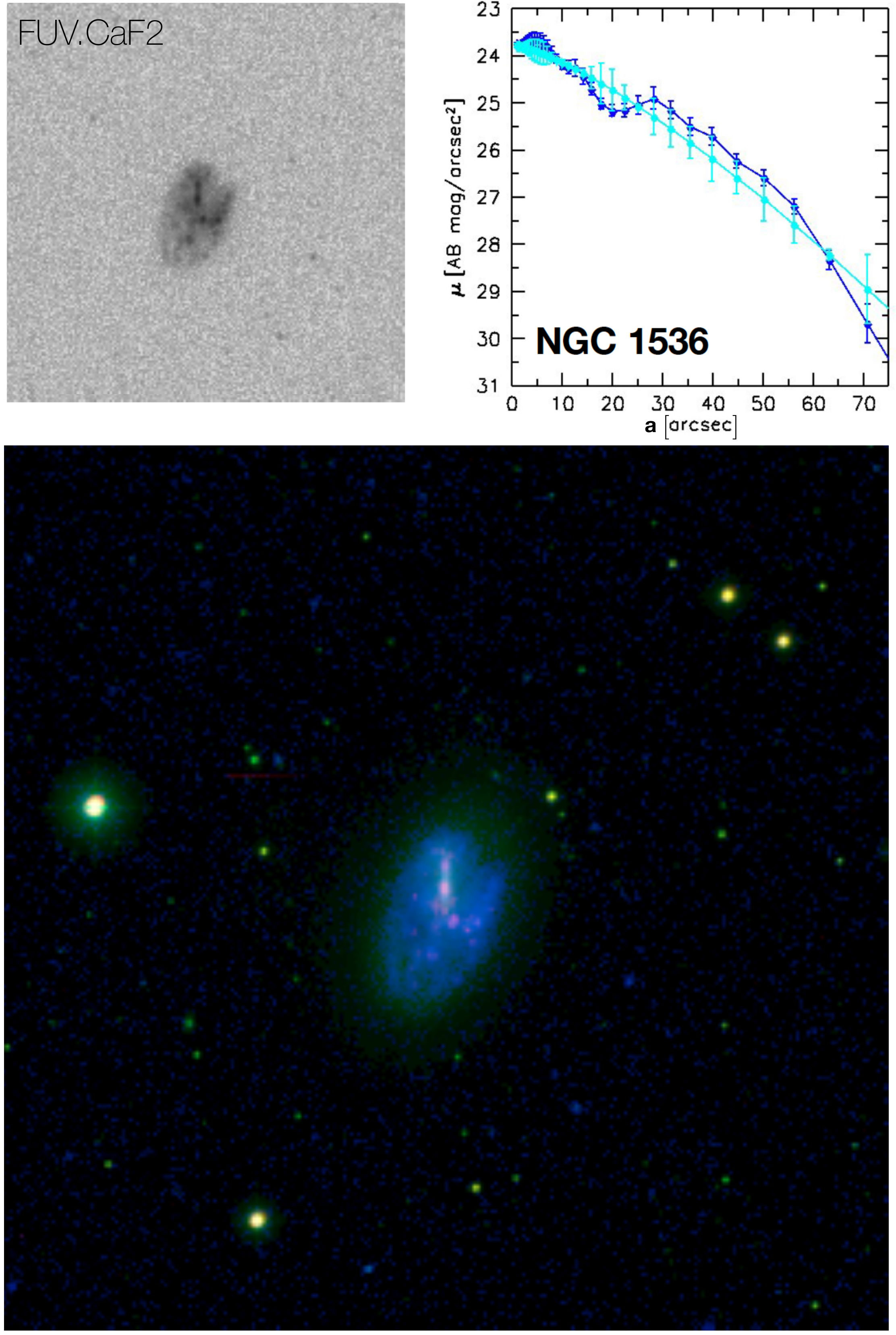}}
	\caption{({\it Top left panel)} {\tt UVIT} FUV.CaF2 image of NGC 1536.  
The image size is 7\arcmin$\times$7\arcmin. 
North is on the top and East to the left. ({\it Top right panel)}
{\tt UVIT} FUV.CaF2 surface brightness profile (blue). The single Sérsic law fit 
is superposed  (cyan) the light profile. ({\it Bottom panel)} Colour composite RGB 
image of NGC 1536 using  \HaN\ and nearby continuum images 
(from Ram2020) as red and green channels and FUV.CaF2 
image as blue channel. Both the \HaN\ and the continuum images
have been re-sampled to the {\tt UVIT} image resolution. 
		\label{NGC1536}}
\end{figure*}

\begin{figure*}
\center
	{\includegraphics[width=13cm]{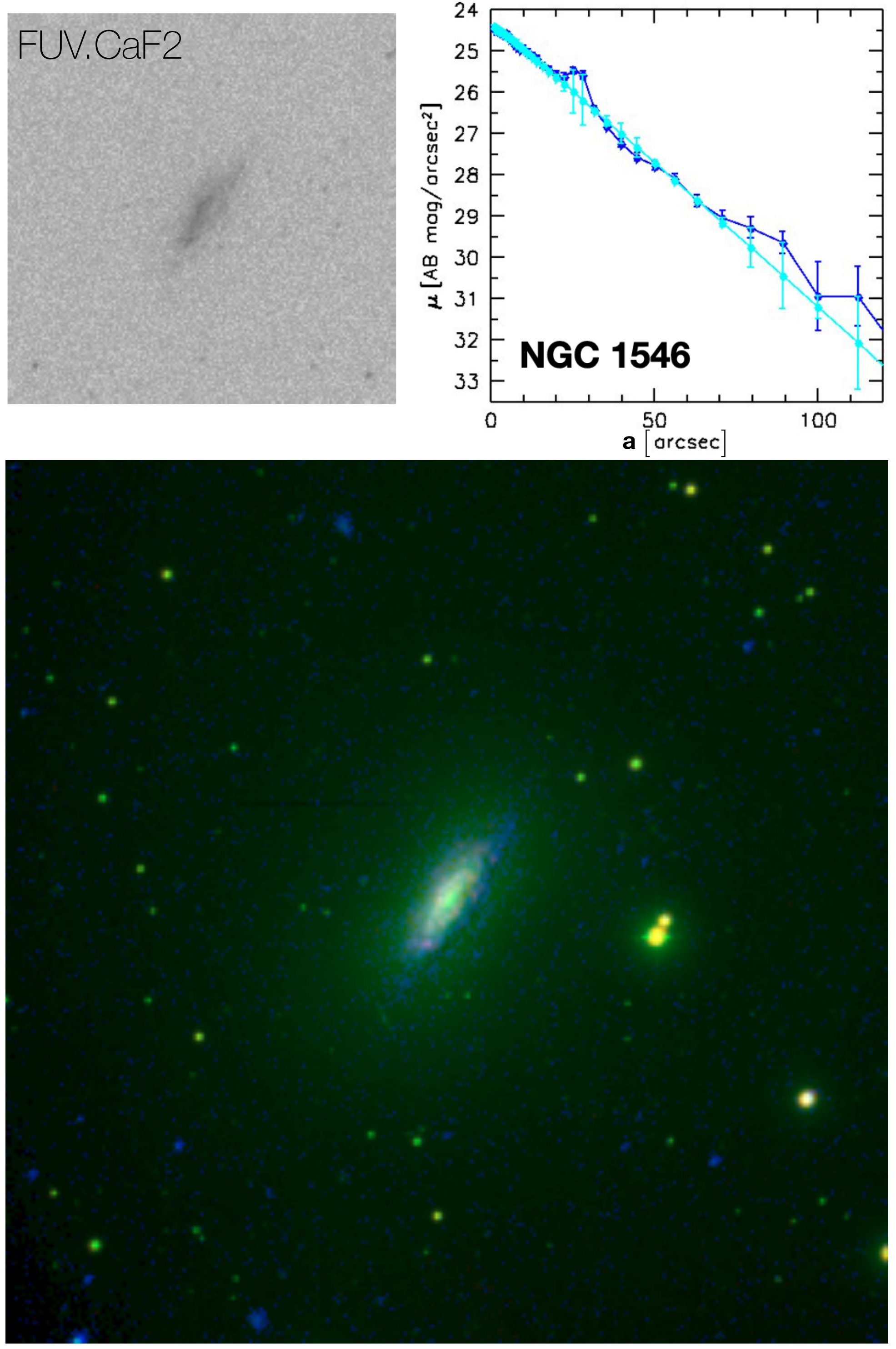}}
	\caption{ As in Figure~\ref{NGC1536} for NGC 1546.
		The image size is 7\arcmin$\times$7\arcmin. North is on the top and East to the left.
		\label{NGC1546}}
\end{figure*}

\begin{figure*}
\center
	{\includegraphics[width=13cm]{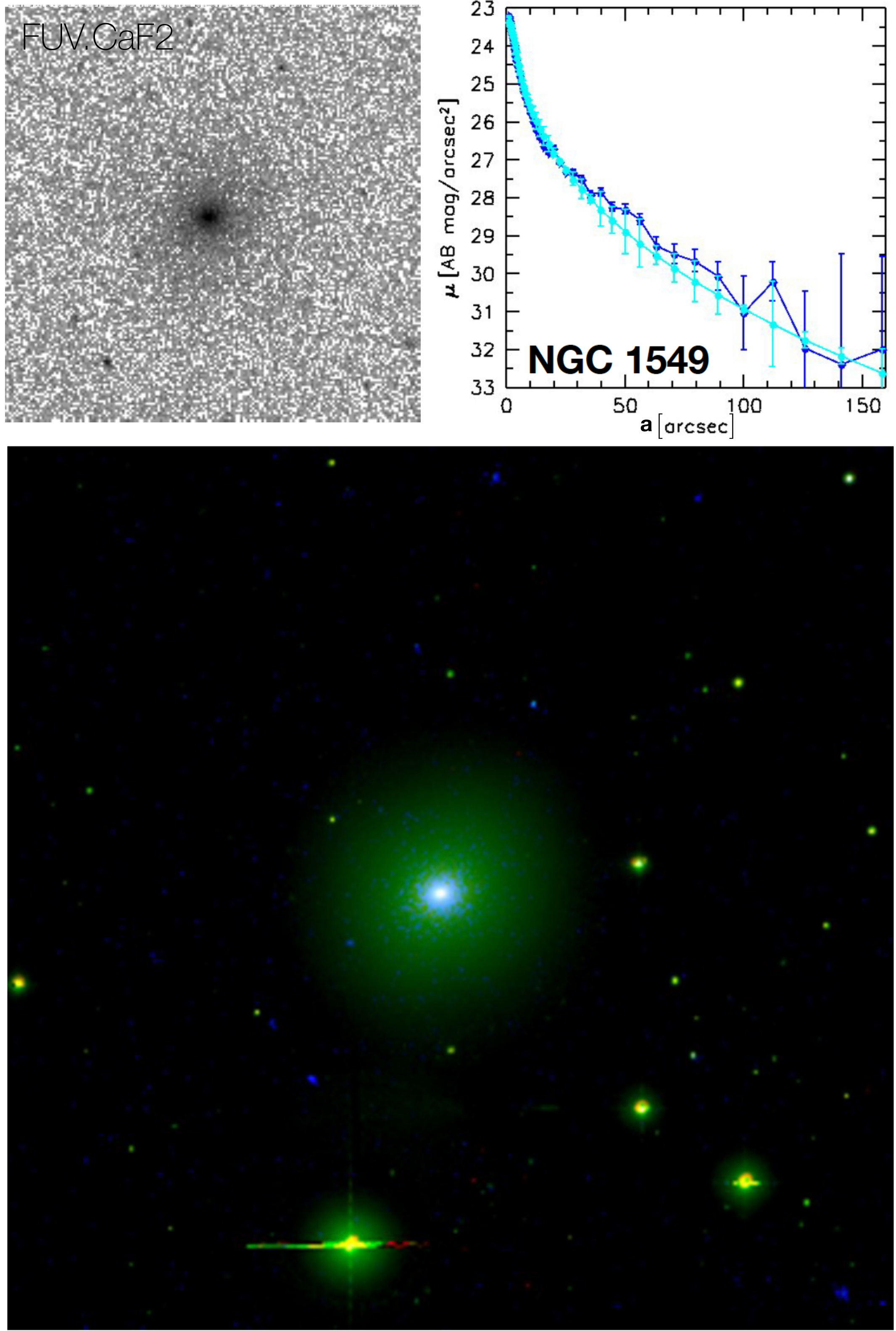}}
	\caption{As in Figure~\ref{NGC1536} for NGC 1549.
		The image size is 8\arcmin$\times$8\arcmin. North is on the top and East to the left.
		\label{NGC1549}}
\end{figure*}

\begin{figure*}
\center
{\includegraphics[width=13cm]{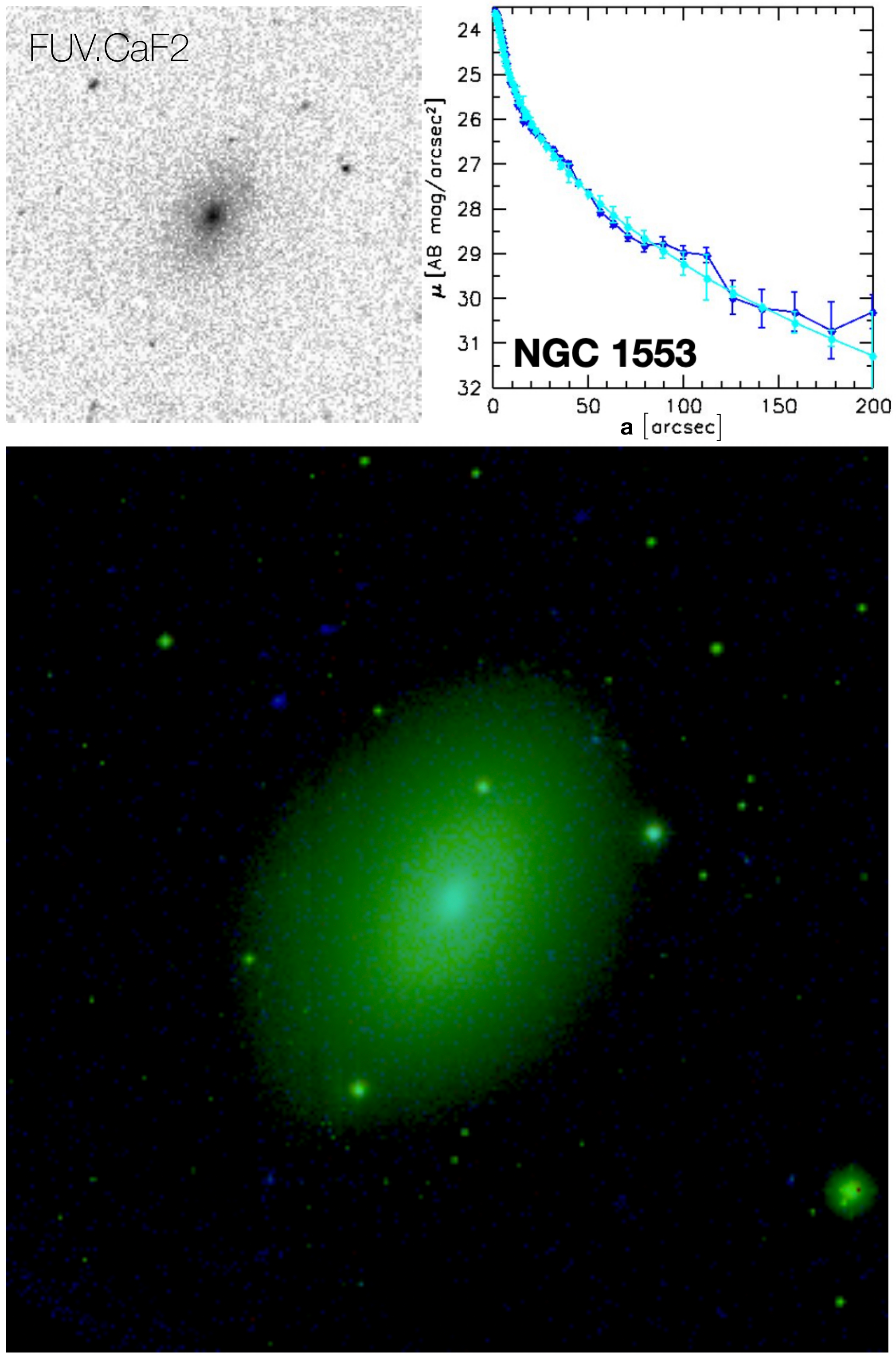}}
	\caption{As in Figure~\ref{NGC1536} for NGC 1553.
		The image size is 7\arcmin$\times$7\arcmin. North is on the top and East to the left.
		\label{NGC1553}}
\end{figure*}

\begin{figure*}
\center
        {\includegraphics[width=18cm]{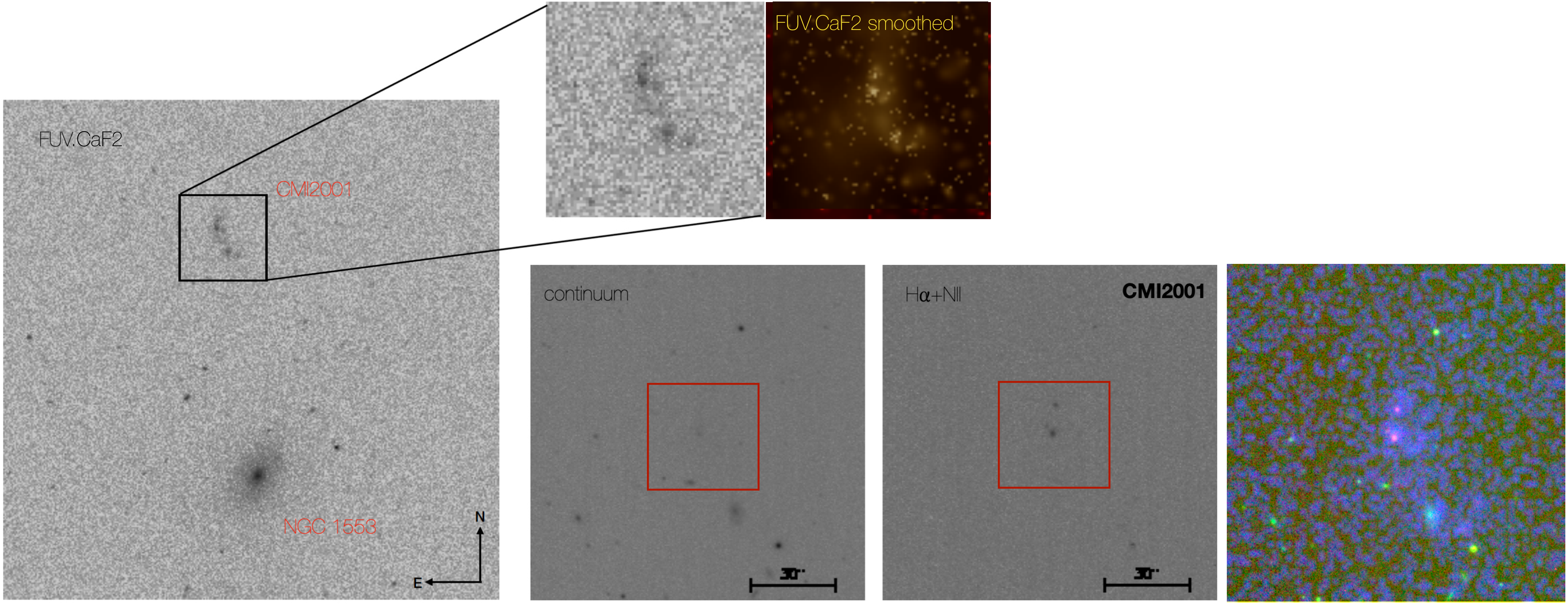}}	
	\caption{ ({\it Left panel)} {\tt UVIT} FUV.CaF2 image of CMI2001 in the 
field of NGC 1553. The size of the figure is 7\arcmin$\times$7\arcmin. 
({\it Top panels}) Zoom on [CMI2001]4136-01 FUV.CaF2 region, 2\arcmin$\times$2\arcmin wide, 
before and after smoothing with {\tt ASMOOTH} and $\tau_{min}$=1.5. 
({\it Bottom panels}) [CMI2001]4136-01 image in  
continuum and \HaN adapted  from Ram2020. The red square encloses the 
CMI2001 region. The bottom right panel shows the RGB colour-composite image 
obtained from the continuum, \HaN, and FUV images used as the green, red and blue 
channels, respectively.}
		\label{CMI2001}
\end{figure*}

\begin{figure*}
\center
	{\includegraphics[width=18.5cm]{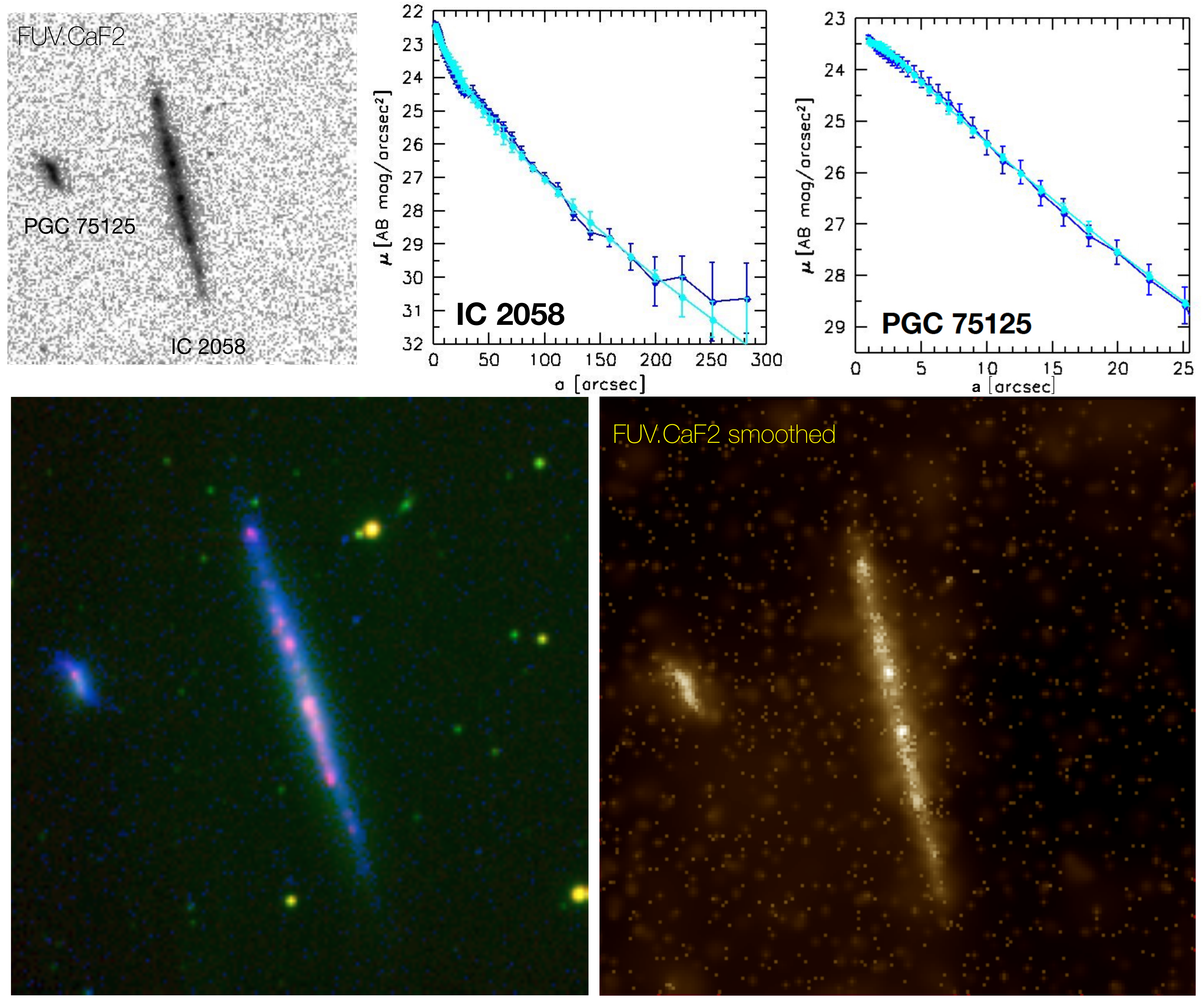}}
\caption{({\it Top left panel)} {\tt UVIT} FUV.CaF2 image of IC 2058 and PGC~75125 
at the centre of the frame  and at the East side, respectively. 
The image size is 5\arcmin$\times$5\arcmin. 
North is on the top and East to the left. {\it (Top central panel)}
FUV.CaF2 surface brightness profile (blue) of IC 2058 and {\it (top right panel)} 
of PGC 75125. Single Sérsic law is superposed (cyan), to fit the light profile.
({\it Bottom left panel})  Colour-composite RGB 
image of IC 2058 and PGC 75125 as in Figure~\ref{NGC1536}. ({\it Bottom right panel}) 
FUV.CaF2 image smoothed using the {\tt AMOOTH} task with $\tau_{min}$=1.5.
	\label{IC2058}}
\end{figure*}

\begin{figure*}
	\center
	{\includegraphics[width=13.8cm]{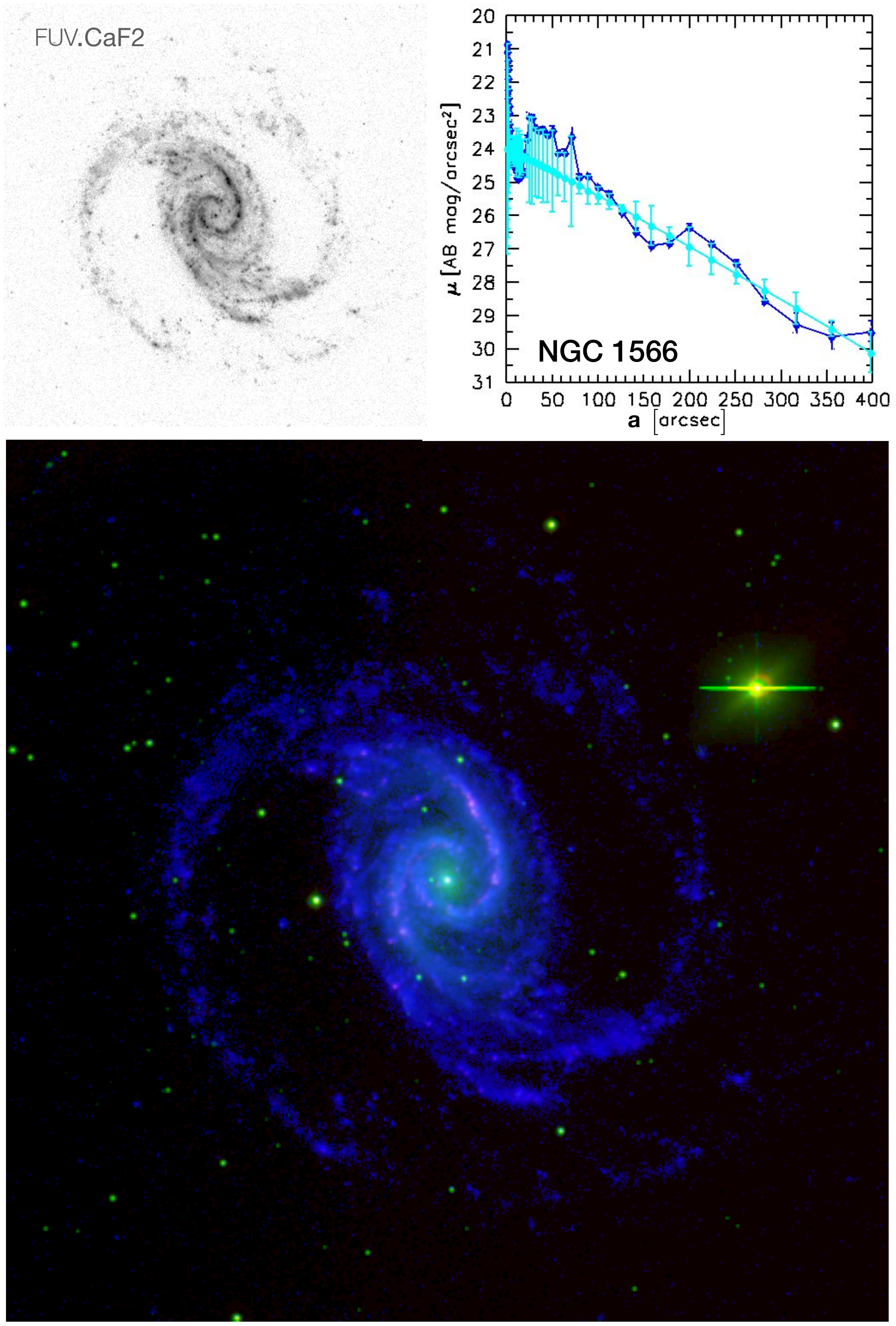}}
	\caption{As in Figure~\ref{NGC1536} for NGC 1566.
		The image size is 13\arcmin$\times$13\arcmin. North is on the top and East to the left.
		\label{NGC1566}}
\end{figure*}

\begin{figure*}
\center
	{\includegraphics[width=17.8cm]{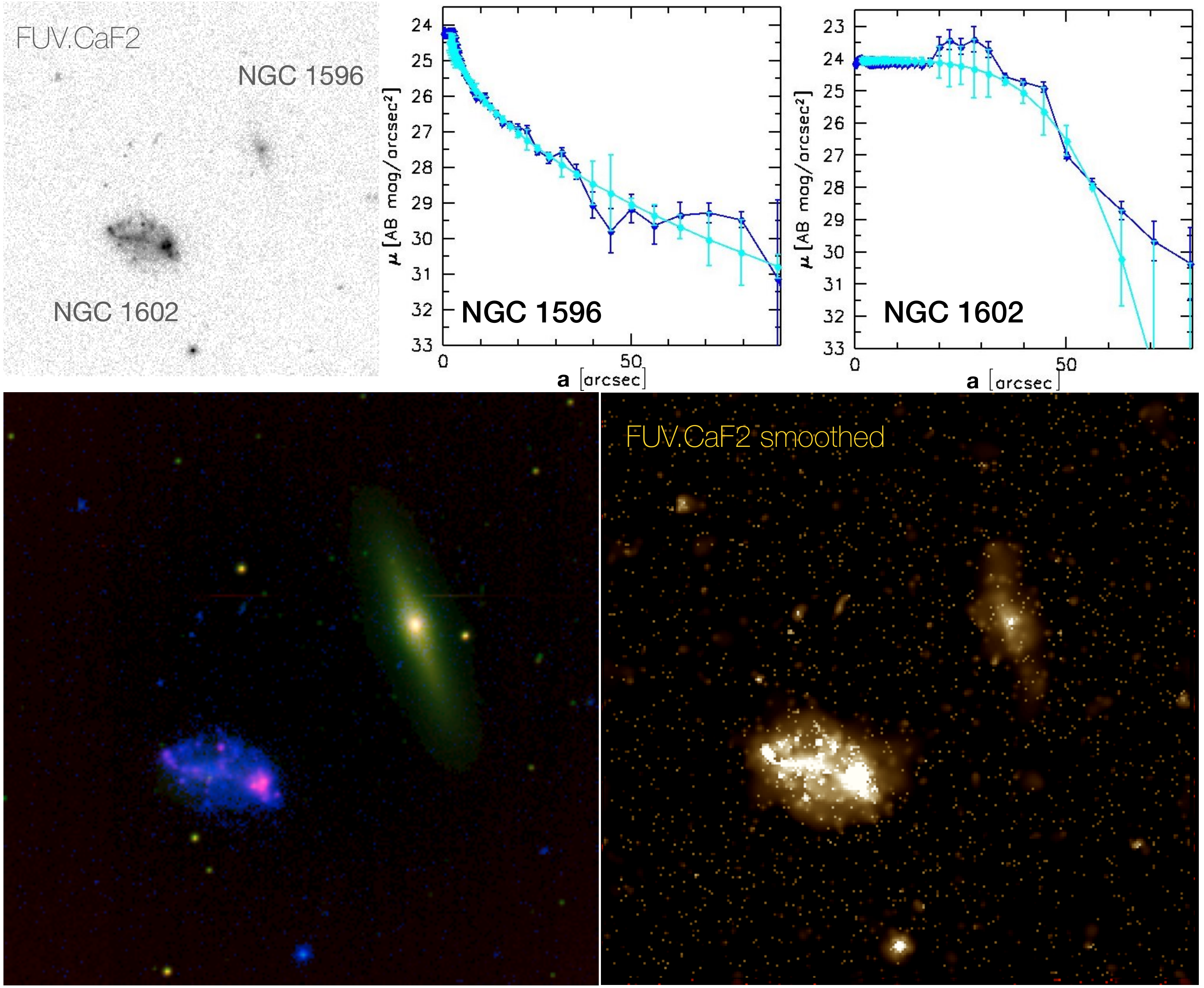}}
	\caption{({\it Top left panel)} {\tt UVIT} FUV.CaF2 image of NGC 1596 (North-west ) 
and NGC 1602 (South-East).
The image size is 7\arcmin$\times$7\arcmin. North is on the top and East to the left.
({\it Top right panels}) FUV.CaF2 surface brightness	
profiles of NGC 1596 and NGC 1602 (blue solid lines). The fit of a 
single S\'ersic law  is overlaid  to the luminosity profile (cyan squares). 
		{\it (Bottom left panel)} 	Colour composite RGB image of NGC 1596
	and NGC 1602 as in Figure~\ref{NGC1536}.({\it Bottom right panel}) 
	FUV.CaF2 image smoothed using {\tt ASMOOTH} task, with $\tau_{min}$=1.5.
		\label{NGC1596}}
\end{figure*}

\subsection{Integrated magnitudes and surface photometry}
\label{surface_photometry}

We derived apparent magnitudes by integrating the surface brightness 
within elliptical isophotes. The {\tt ELLIPSE} task defines the elliptical
contours of the isophotes and accounts for the 
geometrical information contained in the isophotes allowing the 
variation of the ellipticity, $\epsilon=1-b/a$, and position 
angle, $PA$,  along the ellipse major axis, $a$. 
The task provides also a measure of  the
isophotal shape parameter, the so-called $a_4$ parameter from the fourth
cosine component of the Fourier analysis of the fitted ellipse, allowing
to distinguish between boxy ($a_4<0$) and disky ($a_4>$0) isophotes
\citep{Bender1988}. {\tt ELLIPSE}, widely applied to optical images
of ETGs,  has been used by \citet{Jeong2009} and \citet{Marino2011b}
in order to obtained the surface photometry of ETGs from {\tt GALEX} FUV data. 
Since irregular and peculiar features, such as
clumps expected in FUV images, induce sudden variations 
in both $\epsilon$ and $PA$ we notice that the isophotal shape parameter, $a_4$,
looses its physical meaning. For this reason  in Table~\ref{UVIT-results}, which 
collects the relevant parameters derived by our FUV.CaF2 surface photometry  we
provide only the average ellipticity
$\langle \epsilon \rangle$ and position angle $\langle P.A. \rangle$
(columns 4 and 5 respectively).  \\

We estimated magnitude uncertainty by propagating the statistical 
errors on the isophotal intensity. 
Table~\ref{UVIT-results} (col.3) provides the FUV CaF2-1, {\tt UVIT}
integrated magnitudes, corrected for Galactic foreground extinction 
as in Ram2021. The comparison
with the current literature  (Table~\ref{UVIT-sources}) is shown in 
Figure~\ref{comparison_mag} and discussed in Section~\ref{Results}.

FUV.CaF2 images are shown in the top left panel of 
Figures~\ref{NGC1536}, \ref{NGC1546}, \ref{NGC1549}, \ref{NGC1553}, 
\ref{CMI2001}, \ref{IC2058}, \ref{NGC1566}, and \ref{NGC1596}. 
Irregular and peculiar  features are shown by NGC 1536, PGC 75125, 
IC 2058 and CMI[2001]4136-01. Spiral arms are visible in NGC 1546 and NGC 1566.  

The UV surface photometry  of NGC 1553 has been
investigated by \citet{Marino2011b} with {\tt GALEX}.  
Figure~\ref{NGC1553-comparison} 
compares our UVIT  luminosity profile with that of  \citet{Marino2011b}. 
Profiles are consistent.
{\tt GALEX} FUV PSF FWHM is 4.2\arcsec \citep{Morrissey2007}
with respect to the nominal value of 1\farcsec5 of {\tt UVIT} 
\citep{Tandon2017}. We check our PSF using the star TYC 8505-1906-1, 
a single star in the Tycho II stellar catalogue, located in the {\tt UVIT} field D,
which includes NGC 1553.
We fit the stellar profile with a 2D  single Gaussian. The best fit
provides a FWHM$_{RA}$=1\farcsec68 and FWHM$_{Dec}$=1\farcsec47, i.e.
the stellar image has an ellipticity of 0.13. Therefore, the FWHM
value is consistent with nominal value measured by \citet{Tandon2017}.

\subsubsection{{\tt UVIT} FUV.CaF2 versus {\tt GALEX} FUV magnitudes} \label{A1}
\label{magnitudes}
In Figure~\ref{comparison_mag} we compare our  {\tt UVIT} FUV.CaF2
asymptotic magnitudes, obtained integrating the galaxy luminosity 
profiles (see Section~\ref{surface_photometry}) 
with FUV {\tt GALEX} asymptotic magnitudes 
reported in Table~\ref{UVIT-sources}. The comparison includes
Dorado members located in the NGC 1533 sub-structure, 
obtained in the same way by Ram2021.
{\tt UVIT} FUV.CaF2 agree with {\tt GALEX} FUV magnitudes, 
as expected given that the two filters have somewhat similar bandpasses.

\subsubsection{Analysis of the FUV-CaF2-1 surface brightness profiles}
\label{profiles}
\medskip
We fit the azimuthal FUV-CaF2-1 surface brightness profiles of galaxies
with a single S\'ersic law (S\'ersic 1963). We are aware that
in some case it represents a crude representation of the surface 
brightness profile. However, a simple S\'ersic law, provides useful 
information that can be compared with existing results for larger samples. The
S\'ersic law $\mu\propto a^{1/n}$, where $\mu$ is the surface brightness, 
$a$ the semi-major axis and $n$ the S\'ersic index, is a generalization of the 
\citet{Devaucouleurs1992} $a^{1/4}$ and  \citet{Freeman1970} 
exponential laws. The value of  $n$ accounts for  the variety of 
the shapes of the surface brightness profiles  of ETGs with $n=4$ representing 
the `classic' Elliptical's shape. An exponential 
disc \citep{Freeman1970}  has an index $n=1$.
UV surface brightness profiles of ETGs may reach large values of $n$ \citep[see e.g.][]
{Marino2011b}. \citet{Rampazzo2017},  by comparing UV  and optical data
from {\tt Swift-UVOT} of eleven EGTs, suggested that, if $n<3$, the presence
of a disk starts to emerge. The value $n<2.5$  is also adopted 
in optical surveys  to isolate disk galaxies \citep[see e.g.][]{Meert2015}.

The S\'ersic law fit is shown  in  the top right panels of 
Figures~\ref{NGC1536}, \ref{NGC1546}, \ref{NGC1549}, \ref{NGC1553}, \ref{IC2058}, 
\ref{NGC1566} and \ref{NGC1596}, superposed to the {\tt UVIT} FUV.CaF2 surface 
brightness profiles. The fit, that accounts for the {\tt UVIT}-PSF,
is extended to the whole profile up to the background. We did not mask 
FUV bright sub-structures such as  the bar in NGC 1536 and the knots 
in IC2058 or NGC 1602. The values of the S\'ersic index obtained from 
the fits are collected in column 6 of Table~\ref{UVIT-results}. \\

\subsection{The recent Star Formation Rate from FUV.CaF2 integrated galaxy luminosity}
\label{SFR_FUV}

Both \Ha\ and FUV luminosities trace recent SF, although they map slightly different age
ranges. \citet[][their Table~1]{Kennicutt2012} indicated an age range 
of 0--100 Myr and of 0--10 Myr for the stellar populations contributing to
FUV and \Ha, with a mean ages of 10 and 3 Myr, respectively,
although the age strongly depends on stellar metallicity 
\citep[][their Figure 5]{Bianchi2011}. 

From the integrated FUV luminosity of Dorado members we derived
their recent SFR, following \citet[][their equation 3]{Lee2009}:

\begin{equation}
	\label{eq1}
$$	SFR [M_\odot ~yr^{-1}] = 1.4 \times 10^{-28} ~L_{FUV} [erg~s^{-1}Hz^{-1}] $$
\end{equation}

Results are given in column 8 of Table~\ref{UVIT-results}. 
The Galactic extinction correction is 
A$_{FUV}$=7.9$\times$E(B-V) \citep[][see also Ram2021]{Lee2009}
using the E(B-V) values provided  by Ram2020 in Table 3.
No internal dust attenuation has been applied.

\section{Results}\label{Results}

Table~\ref{UVIT-results}  presents results derived from our FUV.CaF2 analysis, 
including luminosity scaled to the distance of 17.69 Mpc. In the following we
discuss the results in terms  of the galaxy type 
classification, which is revised here based on 
the FUV.CaF2 morphology and the S\'ersic index.

\subsection{Individual Notes}
\label{indnotes}

\noindent
\underbar{NGC 1536}~~~~~
The FUV.CaF2 image (Figure~\ref{NGC1536}, top left panel) shows  the irregular 
arm structure of this barred spiral galaxy (SBc, according to {\tt HyperLeda}). 
We can clearly see its off-centered bar.  The average FUV.CaF2 
$\langle PA \rangle=162^\circ$ is consistent with the optical one
 at R$_{25}$ (Table~\ref{UVIT-sources}).
The average ellipticity,  $\langle \epsilon \rangle =  0.40$ (Table~\ref{UVIT-results}), 
differs from the optical one, 0.28 (Table~\ref{UVIT-sources}) 
due to  ellipticity variations of the FUV.CaF2 emission.

The index obtained from the Sersic law fit,  $n=0.76\pm$0.02,
suggests the presence of a disk. 
The \HII\ regions found by Ram2020 stand out in red 
in Figure~\ref{NGC1536}. They are distributed along the  brightest features 
of the FUV.CaF2 emission (bar and arm segments), although the FUV.CaF2 emission 
is more diffuse and extended (see also Section~\ref{Discussion}).

At 30 mag~arcsec$^{-2}$ the  FUV.CaF2 luminosity profiles extends up to 
70\arcsec\ well outside  49\farcs8, the R$_{25}$ isophote in the B-band. 
Notice, however, the larger extension of the R-band continuum (green) with respect
to FUV.CaF2 emission (blue) in the bottom panel of Figure~\ref{NGC1536}.

\medskip

\noindent
\underbar{NGC 1546}~~~~ This galaxy (Figure~\ref{NGC1546}) has un uncertain classification.
\citet{Comeron2014} classified it as E(b)3/(R')SA(r)ab, i.e. it could be
either an ETG, as suggested by {\tt HyperLeda} that reports a classification of S0-a,
or a spiral with an inner ring (r) and an outer pseudo-ring (R'). 
The FUV.CaF2 image (top left panel)
shows diffuse emission in the outer regions  and the colour-composite 
image (bottom panel) reveals a spiral structure embedded 
in a diffuse halo. However, the shape of this galaxy is reminiscent 
of 3D ETGs mentioned  by \citet{Buta2015}, in the R continuum and \HaN\ bands.
The FUV.CaF2 average position angle is consistent with the
optical, 144.8$^\circ$, at R$_{25}$ from {\tt HyperLeda} and
152.4$^\circ$$\pm$0.7 from the Carnegie Galaxy
Survey (CGS hereafter) in the I band \citep{Ho2011}. The average FUV.CaF2 ellipticity, 
$\langle \epsilon \rangle=$0.66, is twice as large as  the optical one 
(Table~\ref{UVIT-sources}).

The fit of the surface brightness profile with a S\'ersic  
index $n=0.98\pm0.04$ is shown in top
right panel of Figure~\ref{NGC1546}. 
The FUV.CaF2 structure results basically  a disk 
with a bump that peaks at about 26\arcsec\, or 2.6 kpc. \citet{Comeron2014} find
a inner closed ring (r) and an outer pseudo-ring (R') in this galaxy. 
The bump we detect in FUV.CaF2 corresponds to the R' pseudo-ring, 
that \citet{Comeron2014} detect at a=28\farcs8.

\medskip
\noindent
\underbar{NGC 1549}~~~~~ The FUV.CaF2 image of this galaxy, shown in Figure~\ref{NGC1549},
contrasts with the large scale optical  system of shells and plumes  
discovered by \citep{Malin1983}. These features, due to old stellar 
populations, are not detected in our red continuum image  (Figure~\ref{NGC1549}, 
green in the composite image) since the short exposure time 
(see also the CGS study by \citet{Ho2011}).
The colour-composite image evidences a blue ring in the inner part 
($\approx$10-20\arcsec radius)  
as revealed from  our FUV.CaF2 luminosity profile  but not from optical data
\citep{Ho2011} and \Ha\ by Ram2020.

The average ellipticity, $\langle \epsilon \rangle$=0.07, and position 
angle,  $\langle PA \rangle$=0$^\circ$ account for the roundness 
and smoothness of the FUV.CaF2 emission unlike
the B-band optical evaluation of {\tt HyperLeda} which provides $\epsilon$=0.15 and  
PA=146.2$^\circ$ at R$_{25}$. The PA of I-band image 
of CGS \citep{Ho2011} 173.8$^\circ$$\pm$2.7 is consistent with
the FUV.CaF2 measure.

Our S\'ersic  fit of the surface brightness profile gives an
index $n=2.86\pm0.28$ suggesting that, in the FUV.CaF2 band,
 NGC 1549 is a E/S0.

\medskip
\noindent \underbar{NGC 1553}~~~~~
The {\tt HyperLEDA} classification for  this galaxy is
S0 with a ring. \citet{Malin1983}  discovered 
a wide system of shells around this galaxy (see their Figure 2). 
The colour-composite image shown in Figure~\ref{NGC1553} does not show 
such features, neither in FUV.CaF2, as expected, nor in the red continuum
image due to the short exposure times.

The recent classification by \citet{Comeron2014} suggests the presence of
resonance rings in this lenticular  SA(rl,nr'l)0$^+$. 
In more detail, they found an 
inner close ring-lens (rl)  with semi-major axis of 35\farcsec4, 
and a nuclear pseudoring-lens structure (nr'l)
with semi-major axis of 9\arcsec \citep[see also][]{Ho2011}. 
Although these features are derived by IR data,  a faint bump appears in the 
FUV.CaF2 luminosity profile of Figure~\ref{NGC1553}
between 30-40\arcsec,  corresponding to  the ring-lens (rl) 
above. 
\Ha\ emission has been detected by Fabry-Perot (FP) high resolution observations 
of the NGC 1553 inner regions \citep{Rampazzo2007}. The FP \Ha\ monochromatic 
map in the inner 30\arcsec\ results quite clumpy
(see Figure 8 in Ram2020).

The average FUV.CaF2 PA and ellipticity are very 
similar to the optical values at R$_{25}$, that are PA=150.4$^\circ$ and $\epsilon$=0.31,
respectively. PA of 151.3$^\circ$$\pm$1.5$^\circ$ is measured in the I-band 
by CGS \citep{Ho2011}.

The right panel of Figure~\ref{NGC1553} shows the S\'ersic fit 
of the surface brightness profile. The S\'ersic 
index obtained,  $n=2.67\pm0.21$, is consistent with that derived by
\citet{Marino2011b}  with the same method using  {\tt GALEX} FUV
observations. This value  suggests that FUV.CaF2 emission is structured in a disk,
i.e. the galaxy is a E/S0 consistently with the {\tt HyperLeda} morphological type
in Table~\ref{UVIT-sources}.

\medskip
\noindent \underbar{[CMI2001]4136-01}~~~~~
This source, detected by Ram2020 (their Figures 9 and 10) 
in \HaN, probing its association to the Dorado group, reveals a complex and 
irregular morphology in our colour-composite  image (Figure~\ref{CMI2001}).  
This source has been classified as a galaxy in the {\tt GALEX}-database.

The bottom panel of Figure~\ref{CMI2001} shows \HII\ regions within 
the faint structure of [CMI2001]4136-01. The case may be reminiscent of \HII\ regions
found by  \citet{Werk2010} in the very periphery of NGC 1533, in the Noth East
sub-structure of Dorado. These seem connected with ongoing interaction
between NGC 1533 and IC~2038/IC~2039 suggested in optical by 
\citet{Cattapan2019}, in \HI\   by \citet{Ryan-Weber2003,Ryan-Weber2004} 
and \citet{Kilborn2009} and in FUV.CaF2  by Ram2021. 
In this context, [CMI2001]4136-01, projecting on an optical shell 
of NGC 1553 (see also Figure 2 in \citet{Malin1983}), could be 
distorted by the interaction with the Dorado centre 
(SCG0414-5559), a region that, at odds with NGC 1533, is \HI\ poor 
\citep{Elagali2019}.

\medskip
\noindent
\underbar{IC 2058 and PGC 75125}~~~~~ IC 2058 and PGC 75125 
are part of the Dorado central compact group SCG 0414 -5559 \citep{Iovino2002}. 
A distortion of PGC 75125 is evident from the FUV.CaF2 image (Figure \ref{IC2058}).
The two galaxies, separated by $\Delta V_{hel}=28$ \kms,  are 
likely interacting, as suggested by both the shape of the present FUV.CaF2
emission and \HI\ observations by \citet[][their Figure 14]{Elagali2019}.

The FUV.CaF2 image of IC 2058, seen edge-on,   
does not show distortions. The smoothed image, shown in the
bottom right panel of Figure~\ref{IC2058}, shows the
presence of some extra-planar features, not
detected in \Ha\ both by \citep[][]{Rossa2003} 
and Ram2020.

The geometrical parameters of IC 2058, PA=17.9$^\circ$
and $\epsilon$=0.85, at R$_{25}$ in Table~\ref{UVIT-sources}
are very similar to the average  $\langle \epsilon \rangle=$0.74 and 
$\langle PA \rangle=18^\circ$ obtained from the 
FUV.CaF2 surface photometry (Table~\ref{UVIT-results}).

The single S\'ersic law fits of both galaxies luminosity profiles,
shown in the top central and right panels of Figure~\ref{IC2058},
with index $n=1.48\pm0.10$ for IC 2058 and $n=1.20\pm0.03$ for PGC 75125
trace the presence of a disk in both these galaxies (Table~\ref{UVIT-results}).

\medskip
\noindent
\underbar{NGC 1566}~~~~~ The FUV.CaF2 emission (Figure~\ref{NGC1566}) of
 this grand design spiral, classified by \citet{Comeron2014} 
 (R')SAB(r'l,s,nb)b, is very extended with the outermost spiral 
 arm  out to about 6\farcmin7 (about 31 kpc). This radius is larger
than both the optical B-band radius at $\mu_{25}$, 3\farcmin62
({\tt HyperLeda}), and the I-band emission radius measured at 26.5 mag~arcsec$^{-2}$
that reaches 4\farcmin06 \citep{Ho2011} (see Section~\ref{Discussion}
 and Table~\ref{UVIT-opt}.)

The optical position angle and ellipticity measured at R$_{25}$, 
PA=44.2$^\circ$ and $\epsilon$=0.32, 
differ from those provided by the FUV.CaF2 
surface photometry. FUV.CaF2 isophotes appear nearly round 
with average ellipticity 0.05  and position angle 15$^\circ$.
The PA measured in the I-band by CGS is 26.4$^\circ$$\pm$4.7$^\circ$.

In the top right panel of Figure~\ref{NGC1566} are shown our
surface brightness profile and single S\'ersic law fit. The fit does not
interpret the inner ring.
The S\'ersic index, $n=0.94\pm0.05$, follows the general disk structure, 
while the actual profile shows deviations corresponding to the spiral arms. 

\medskip
\noindent
\underbar{NGC 1596 and NGC 1602}~~~~~

Our UVIT image is the first UV view of both these systems. Our composite image 
(Figure~\ref{NGC1596}) shows that the FUV.CaF2 emission dominates the light 
of the irregular galaxy NGC 1602 whereas it is mixed to  older stellar 
populations in NGC 1596. 
We enhanced the signal of faint structures using {\tt ASMOOTH} task
to look for a FUV connection between the two galaxies following 
the behaviour the \HI\ emission \citep{Bureau2006,Elagali2019} 
without success.  

The smoothed image (bottom right panel), 
however, reveals extra-planar light, so the FUV disk  looses 
the regular shape shown in the optical bands (bottom left panel).  

The value of the S\'ersic index in Figure~\ref{NGC1596} (top: central and left 
panels) $n=2.12\pm0.18$  (Table \ref{UVIT-results}) is  consistent 
 with the presence of a disk in NGC 1596. The average position angle and 
ellipticity, 0.5 and 20$^\circ$, are consistent with 
{\tt HyperLeda} values (Table~\ref{UVIT-sources}) and the I-band measure of the 
PA 20.0$^\circ$$\pm$0.8$^\circ$ from CGS.
 
The surface brightness profile of NGC 1602 is very irregular as the shape
of the galaxy which is classified a SBm in optical. 
The average  ellipticity and position angle, 0.3 and 80$^\circ$
are consistent with {\tt HyperLeda} values (Table~\ref{UVIT-sources}). 

The S\'ersic index, $n=0.25\pm0.04$, 
results from the irregular shape of the FUV.CaF2 light distribution in this galaxy. 
Notice that the case $n=0.5$ corresponds to a Gaussian.

\begin{center}
	\begin{table*}[t]%
		\centering
		\caption{Integrated magnitudes of Dorado in UVIT FUV.CaF2. 
			\label{UVIT-results}}%
		\begin{tabular*}{40pc}{@{\extracolsep\fill}lccccccc@{\extracolsep\fill}}%
			\hline
			\textbf{Field} &  \textbf{ID}     & \textbf{FUV.CaF2}  & $\langle \epsilon \rangle$ &  $\langle PA \rangle$ & n & L$_{FUV.CaF2}$ & SFR\\
			\textbf{ }     &  \textbf{source} & \textbf{AB [mag]} &        & \textbf{[deg]}  &  & \textbf{10$^{26}$ [erg s$^{-1}$ Hz$^{-1}$]} & \textbf{[M$_\odot$ yr$^{-1}]$}\\
			\hline
			A &  {\bf NGC 1536}  & 15.90$\pm$0.02   &  0.40 & 160  & 0.76$\pm$0.02& 5.95$\pm$0.05 & 0.083$\pm$0.001     \\
			\hline
			B   & {\bf NGC 1546} & 17.24$\pm$0.11&0.66 & 141  & 0.98$\pm$0.04& 1.74$\pm$0.08&  0.024$\pm$0.001     \\
			\hline
			C    & {\bf NGC1549} & 16.92$\pm$0.11 &  0.07   & 0 & 2.86$\pm$0.28 &2.32$\pm$0.01 & 0.033$\pm$0.001 \\
			&  [CMI2001]4136-01 & 19.13$\pm$0.34  & \dots  & \dots & \dots & 0.30$\pm$0.04 & 0.004$\pm$0.001\\
			\hline
			D    &  {\bf NGC 1553}         & 16.44$\pm$0.13  &   0.36 & 150   &2.67$\pm$0.21 &3.63$\pm$0.20 & 0.051$\pm$0.003\\
			\hline
			E  & {\bf  IC 2058}     & 16.22$\pm$0.04 &  0.74 & 18 & 1.48$\pm$0.10& 4.43$\pm$0.07& 0.062$\pm$0.001 \\ 
			& {\bf PGC 075125}  & 18.07$\pm$0.10      & 0.36 & 20 & 1.20$\pm$0.03& 0.81$\pm$0.03& 0.011$\pm$0.0005\\
			\hline
			F   & {\bf NGC 1566} & 12.13$\pm$0.03 & 0.05 & 15 & 0.94$\pm$0.05 & 175.30$\pm$1.93 & 2.455$\pm$0.027 \\       
			\hline
			G   & {\bf NGC 1596}  & 17.96$\pm$0.21  &  0.52      & 20 & 2.12$\pm$0.18& 0.95$\pm$0.08 &  0.013$\pm$0.001\\ 
			& {\bf NGC 1602} &  15.06$\pm$0.04  &   0.32     & 80 & 0.25$\pm$0.04&12.94$\pm$0.21 & 0.181$\pm$0.003 \\          
			\hline
		\end{tabular*}
		\tablefoot{Col. 2 gives the galaxy identification;  
			Col. 3 gives the extinction corrected FUV integrated magnitude; Col. 4 gives the average ellipticity;
			Col. 5 gives the average Position Angle; Col. 6 gives the Sers\'ic index; Col. 7 gives the total 
			absolute FUV luminosity; col. 8. gives the SFR as derived from eq.1.}
	\end{table*}
\end{center}

\bigskip
We summarize the results  of our   FUV.CaF2 morphological and structural analysis.

\begin{itemize} 
	
	\item{All surface brightness  profiles are fitted by a 1D single S\'ersic law
with $n<3$ suggesting that the  FUV.CaF2 structure of most of the 
Dorado members we analysed, including ETGs, has a disk shape.
Such FUV structures have been formed via dissipative mechanisms 
\citep{Rampazzo2017}.}
			
\item{NGC 1536 shows  an off-centred bar. 
FUV.CaF2 rings are frequently detected both among ETGs and LTGs. 
Our findings support previous {\tt GALEX} observations,
\citep{Jeong2009,Marino2011a, Marino2011b} showing that these rings 
are SF sites \citep{Bianchi2011}, associated to \HII\ regions, 
as for NGC 1546, NGC 1566, NGC 1581  (Ram2020) and NGC 1533 
(Ram2021).}
		
\item{We observe FUV.CaF2 extra-planar emission from the disk of IC2058 and NGC 1596. 
(Figure~\ref{IC2058} and Figure~\ref{NGC1596}, right bottom panels).}
We fail in revealing FUV.CaF2 diffuse emission
in correspondence to \HI\ emission in
between IC 2058/PGC 75125 and NGC 1596/NGC 1602
\citep{Chung2006,Elagali2019}. Further insights on this point coming
from deep optical images are discussed in the next section.

\item{Asymmetries are visible in NGC 1536, PGC 75125, NGC 1566 and CMI[2001]4136-01
likely due to tidal perturbations.}

\end{itemize}

\section{FUV.CaF2 versus deep optical imaging: looking for XUV emission}
\label{Discussion}

We intend to test the presence of XUV disks
in the Dorado members, i.e. the presence of FUV emission 
at large galacto-centric distances, up to several
optical radii. Roughly 30\% of  spiral galaxies 
in the Local Universe show XUV features 
\citep{Thilker2007,Gil2008,Thilker2010}. This
phenomenon is relevant for understanding how FUV observations 
intercept faint SF levels \citep[see e.g.][]{Bigiel2010}.
For comparison, we consider also the FUV.CaF2 extension 
versus optical in ETGs.

The FUV-optical comparison performed
in Section~\ref{Results}, made on the basis of continuum observations 
from the \HaN\ data-set of  Ram2020, is hampered by the short
exposure time through the narrow band optical filters used.
In this section we compare the FUV.CaF2 extension of our targets 
with optical radius at 29 and 28 mag~arcsec$^{-2}$ in 
$g$ and $r$ bands, respectively, as results from
the deep photometry by Ragusa et al. (2022, in preparation). 
Table~\ref{UVIT-opt} reports the above values together with
the R$_{26.5}$ radius in I-band measured by \citet{Ho2011}.

\begin{figure*}
	\center
{\includegraphics[width=18.5cm]{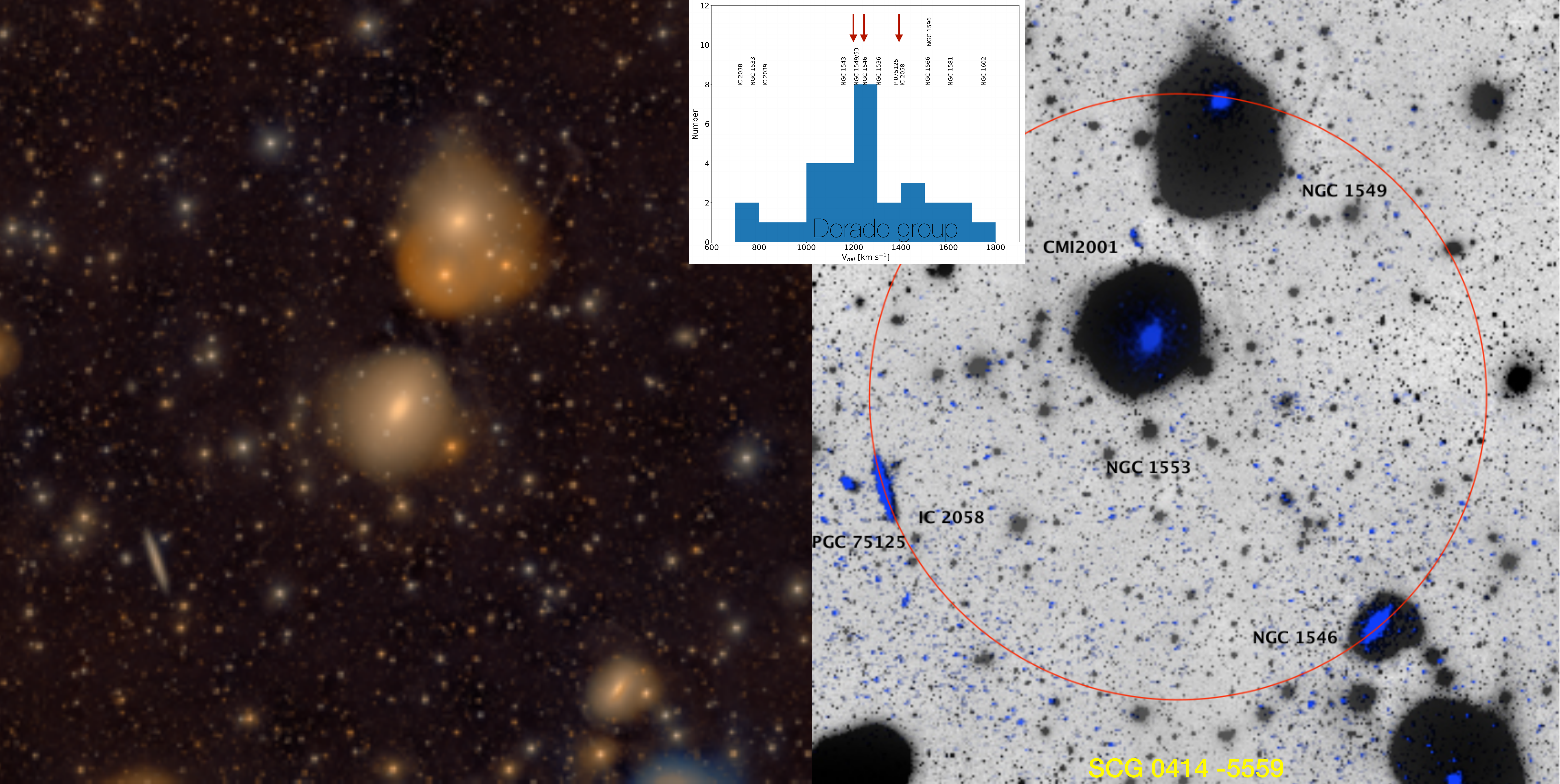}}
\caption{The compact group SCG 0414 -5559. The left  panel 
shows the colour composite VST image of the Dorado group centre (North is on the top 
and East to the left). The field of view  is 51\farcmin77$\times$47\farcmin52.
The SCG is at the centre of the Dorado members velocity distribution 
in the \citet{Kourkchi2017} group definition (central panel). 
The colour RGB image is created adopting the \citet{Lupton2004} scheme, 
using the $r$-band image as R channel input, the $g$-band as B channel input, 
and the average image between $r$ and $g$ bands as G channel input. 
Notice faint shell/ripples structures in NGC 1549 and NGC 1553 
\citep[see][]{Malin1983} and in the south-east side
of NGC 1546. In the right panel an unsharp mask of the  SCG is shown.
In red is drawn the  circle that encloses the compact group,
SCG 0414-5559, identified by \citet[][see her Figure 3 and Table~1]{Iovino2002}.  
In addition to galaxies forming the compact group we labelled PGC 75125, a 
physical companion of IC 2058 \citep[see e.g.][]{Elagali2019} and CMI2001,
not present in \citet{Iovino2002}.
Our B, C, D, E {\tt UVIT} fields are superposed to the unsharp masking 
of the $g$-band VST image in order to show  extension and distribution 
of the FUV emission within the compact group members. 
}
\label{Dorado-group-barycentre}
\end{figure*}

\begin{figure*}
	\center
	{\includegraphics[width=18.5cm]{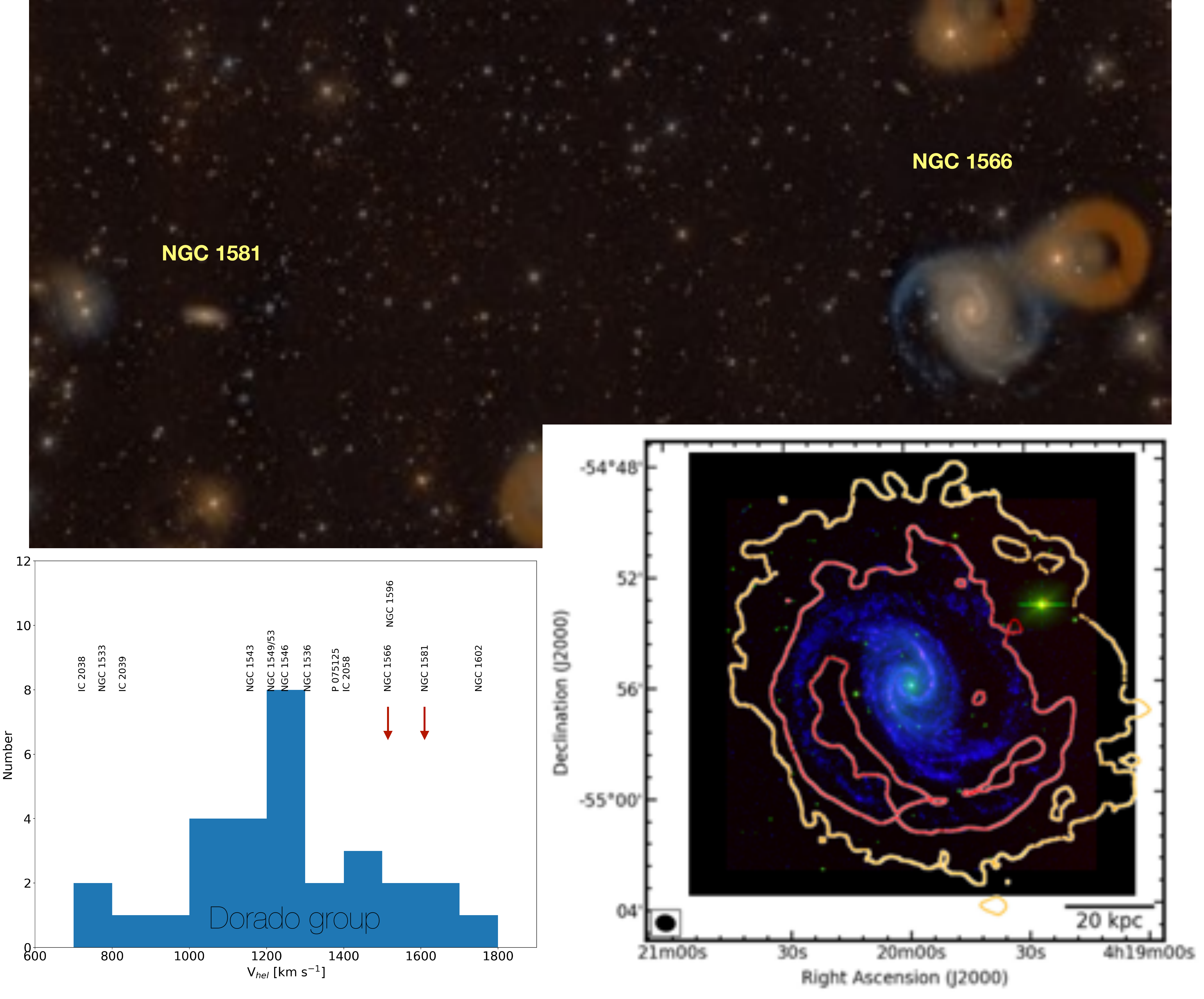}} 
	\caption{NGC 1566 and NGC 1581. 
In the top panel is shown the VST image colour composite image (RGB as in 
Figure~\ref{Dorado-group-barycentre}) of NGC 1566 and NGC 1581 members 
(North is on the top and East to the left, FoV=60\farcmin80$\times$31\farcmin37).
Their projected separation  is 40.84\arcmin, while they are separated by 96 \kms,
as shown in the bottom left panel. The bottom right panel shows that the FUV emission 
of NGC 1566 (13\arcmin$\times$13\arcmin see Figure~\ref{NGC1566}) extends up to 
the outer columns density of \HI\ at 3.7$\times$10$^{20}$ cm$^{2}$ (red) and, 
especially in the North region, almost to 0.6$\times$10$^{20}$ 
cm$^{2}$ (orange)  reported by \citet{Elagali2019}. 	}	
		\label{UVIT_VST_NGC1566_view}
\end{figure*}

\begin{figure*}
	\center
	{\includegraphics[width=18.5cm]{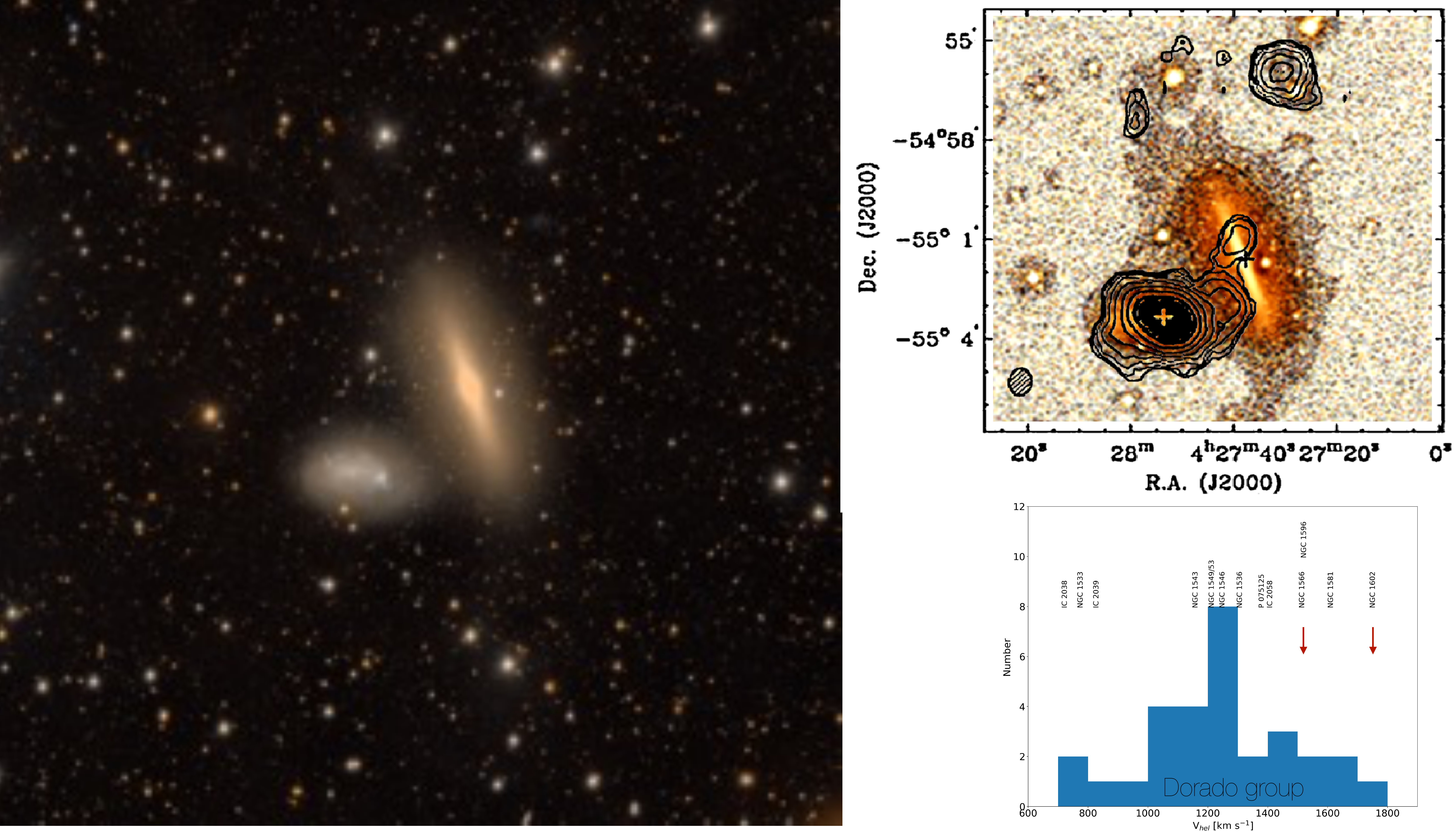}} 
	\caption{The physical pair NGC 1596/NGC 1602. In the redshift space
		these galaxies are seprated by 230 \kms, as shown by the bottom right panel.		
		The VST colour composite image  (RGB as in Figure~\ref{Dorado-group-barycentre})
		of the pair (left panel, North on is
		the top and East to the left, FoV=17\farcmin34$\times$17\farcmin) shows that 
		NGC 1602 outskirts appear distorted 
		and, in projection, roughly in contact with NGC 1596, as well as 
		 \HI\ contour levels  (4, 6, 9, 13.5, 20.3, 30.5, 45.7, 68.6, 
		102.8 and 154.2 $\times$ 10$^{19}$ cm$^{-2}$) by \citet{Chung2006}
		shown in the top right panel. These contours are superposed to our VST
		unsharp masked  image  in the top right panel.
		Crosses indicate the optical centres of the two galaxies, and the synthesized 
		beams (49.9$\times$37.6 arcsec$^2$) are also shown at the bottom left of this panel.	
		Similar results for \HI\ are  in Figure~14 of \citet{Elagali2019}. 
		\label{UVIT_HI_NGC1596_view}}
\end{figure*}

\subsubsection{The Dorado core}
\label{Dorado-compact}

The left panel of  Figure~\ref{Dorado-group-barycentre} shows 
a {\tt VST} colour composite image of the compact group.
The image includes NGC 1546, NGC 1549 NGC 1553 and IC 2058, 
observed with {\tt UVIT} FUV.CaF2 fields B, C, D, E 
(see Table~\ref{UVIT-sources} and Figure~\ref{Dorado_map}). 
The  {\tt VST} images include also PGC 75125 and of 
[CMI2001]4136-01 projected between NGC 1549 and NGC 1553.
An unsharp masking of the same field, in the $g$-band, is shown in the right
panel. We notice that NGC 1546, in addition to NGC 1549 and NGC 1553
\citep{Malin1983}, shows a faint shell/ripple in its South-East side. 
Shells are mostly revealed in ETGs 
while they are very rare, if any, in LTGs \citep[see e.g.][]{Schweizer1992}.  
The presence of shells/ripples plus \HII\ regions 
in the central spiral like structure of NGC 1546 (Ram2020) 
make this galaxy a  borderline object between ETGs and LTGs. 
The morphological type (T=-0.4) and the classification by \citet{Comeron2014}
is appropriated.

The radial extension on the {\tt VST} image of NGC 1546, NGC 1549,
NGC 1553 is  larger  than FUV.CaF2 emission radius by 2.3, 3.2 and 
3.4 times in $g$ and by 2.1, 3.5 and 3.5 in $r$ band (Table~\ref{UVIT-opt}).
At odds, the FUV.CaF2 emission of IC 2058 has a radius of 3\farcmin2 
comparable to the galaxy $g$ band extension (1.1 times) but larger 
than $r$ band extension which is 0.7 times the FUV ones.

At the light of the {\tt VST} image, the structure shown by PGC 75125, 
the physical companion of IC 2058, in FUV.CaF2 (Figure~\ref{IC2058}) 
reveals to be either a warped disk or a bar-like structure which is 
embedded in a spheroidal halo. The optical radii reported in Table 
(Table~\ref{UVIT-opt}) are 0\farcmin62 and 0.57 in $g$ and $r$ bands, respectively,
and 0\farcmin42 in  FUV.CaF2 at $\mu_{FUV.CaF2}=29$ mag arcsec$^{-2}$,
i.e. the extension in optical PGC 75125 is 1.5 times larger than
in FUV.CaF2. We conclude that the compact group members do
not show signature of XUV features.

\begin{center}
	\begin{table*}
		\centering
		\caption{Radial extension of targets in {\tt FUV.CaF2}, I, $g$ and $r$ bands \label{UVIT-opt}}%
		\tabcolsep=0pt%
		\begin{tabular*}{30pc}{@{\extracolsep\fill}lccccc@{\extracolsep\fill}}
			\hline
			\textbf{Galaxy} & \textbf{{\tt FUV.CaF2} R$_{30}$}  & \textbf{CGS R$_{26.5}$} & \textbf{VST R$_{29g}$}  & \textbf{VST R$_{28r}$}  \\
			\textbf{}       &  \textbf{[arcmin]}  &  \textbf{[arcmin]}     & \textbf{[arcmin]}            &   \textbf{[arcmin]}  \\
			\hline
IC 2038   & 1.5$^{1}$&  \dots  &  1.1$^{3*}$  & 1.0$^{3*}$ \\
IC 2039   & 0.3$^{1}$& \dots   &  1.0$^{3*}$  & 0.6$^{3*}$  \\
NGC 1533  & 1.5$^{1}$ &1.63   &    5.2$^{3}$ &  4.3$^{3}$    \\
NGC 1546  & 1.7           & 2.81   &   4.0$^{4}$ & 3.5$^{4}$ \\
NGC 1549  & 2.0           & 2.65   &   6.5$^{4}$ & 7.1$^{4}$ \\
NGC 1553  & 2.0           & 7.23   &  6.7$^{4}$& 6.2$^{4}$ \\
IC 2058   & 3.2            & \dots  &  3.4$^{4}$ & 2.3$^{4}$ \\
PGC 75125 & 0.4$^{2}$            & \dots  & 0.6$^{4}$ & 0.6$^{4}$ \\
NGC 1566  & 6.7             & 4.06   &  8.0$^{4}$ & 6.5$^{4}$ \\
NGC 1596  & 1.3             & 3.02   &  6.0$^{4}$ & 5.5$^{4}$ \\
NGC 1602  & 0.8             & \dots  &  3.6$^{4}$ & 3.6$^{4}$ \\			\hline
		\end{tabular*}
		\tablefoot{Col. 1 gives the galaxy identification;  Col. 2 reports the
semi-major axis R$_{30}$ i.e. the extension of the galaxy at 
$\mu_{FUV.CaF2}$=30 mag arcsec$^{-2}$ measured along the 
FUV.CaF2 luminosity profile. Correspondingly col. 3  provides R$_{26.5}$, the radius at 
26.5 mag arcesec$^{-2}$, derived from I-band by CGS \citep{Ho2011}; 
Col.s 4  and 5 give R$_{29g}$ and R$_{28r}$, the semi-major axis 
the galaxy luminosity profile at 29 and 28 mag arcesec$^{-2}$ in the $g$ and $r$ bands, 
respectively. Notes:$^{1}$ data are from Ram2021; $^{2}$ the radius for 
PGC 75125 refers to $\mu_{FUV.CaF2}$=29 mag arcsec$^{-2}$;  $^{3}$ data are from
\citet{Cattapan2019}; $^{4}$ data are from Ragusa et al. (2022 in preparation). 
$^*$ The radii for IC 2038 and IC 2039 in colums 5 and 6 refer to 
$\mu_{g}$=25 mag~arcsec$^{-2}$ and $\mu_{r}$=24 mag~arcsec$^{-2}$,
respectively.
		}
	\end{table*}
\end{center}

\noindent 
\subsubsection{NGC 1566 and NGC 1581}

The top panel of Figure~\ref{UVIT_VST_NGC1566_view} shows the {\tt VST} 
colour composite image  of NGC 1566 and NGC 1581. 
The projected separation between  these galaxies, 210\,kpc, is quite
large although the recession velocity separation, $\Delta V_{hel} =$ 96, is
 \kms (bottom left panel of Figure~\ref{UVIT_VST_NGC1566_view})
compatible with galaxies being associated.

The {\tt UVIT} field F includes NGC 1566. 
NGC 1581 was not observed neither by {\tt GALEX} nor by {\tt UVIT}.
The grand design spiral NGC 1566 dominates the FUV emission of the Dorado group 
(Table~\ref{UVIT-results}, see also Figure~\ref{comparison_mag}). It 
extends up to 1.65$\times$ R$_{26.5}$  and 0.83$\times$ 
R$_{29g}$ and is similar to R$_{28r}$ (Table~\ref{UVIT-opt}).
The bottom right panel of Figure~\ref{UVIT_VST_NGC1566_view} 
shows that the FUV.CaF2 emission  extends almost up to 
the \HI\ column density of 3.7$\times$10$^{20}$ cm$^{2}$, 
marked by the red contours. 

\noindent 
\subsubsection{NGC 1596 and NGC 1602}

The radial velocities of NGC 1596 and NGC 1602 differ by 
230 \kms\ as shown by the red arrows in the bottom right panel of 
Figure~\ref{UVIT_HI_NGC1596_view}.The left panel of this Figure shows 
a colour-composite image of the pair. NGC~1602 is enclosed in a 
halo stretched towards the NGC 1596 halo. 

{\tt UVIT} FUV.CaF2 observations (Figure~\ref{NGC1596}) show that
the two galaxies are sharply separated and without any outer halo.
However, the extra-planar features shown by adaptive smoothing applied to the
FUV.CaF2 image (bottom right panel of Figure~\ref{NGC1596})
may trace the \HI\ connection  \citep{Chung2006} between 
the two galaxies shown in the top left panel of 
Figure~\ref{UVIT_HI_NGC1596_view} in which the neutral gas contour are overlaid. 
The on-going interaction between these two Dorado members 
distorts their halos as evidenced by the unsharp masking of the
{\tt VST} image. 

The radius is about 1\arcmin3 at $\mu_{FUV.CaF2}=30$ mag~arcsec$^{-2}$
for NGC 1596. The optical emission from {\tt VST} of NGC 1596 
is 5\farcmin5 at $\mu_{g}=29$ mag~arcsec$^{-2}$ and 6\arcmin\
at  $\mu_{r}=28$ mag~arcsec$^{-2}$ indicating that the optical emission is 
4.6 and 4.23 times larger than the FUV.CaF2 emission.
NGC 1602 at $\mu_{FUV.CaF2}=30$ mag~arcsec$^{-2}$ the radius of
0\farcmin8, i.e 4.5 times smaller than both the R$_{29g}$ and R$_{28r}$ radii from 
{\tt VST} frames (see Table~\ref{UVIT-opt}). None of these galaxies
host a XUV disk.

\begin{figure*}
	{\includegraphics[width=9.cm]{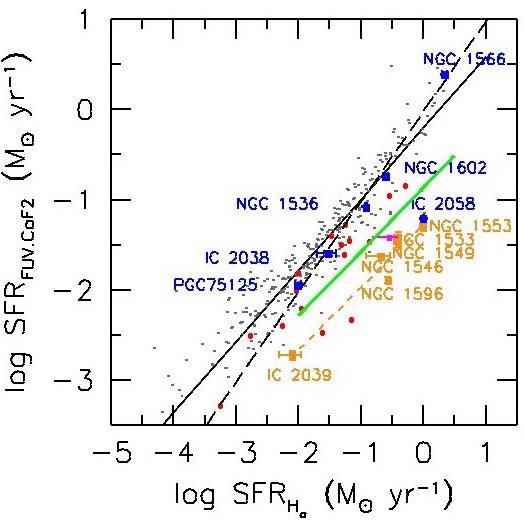}}
	{\includegraphics[width=9.4cm]{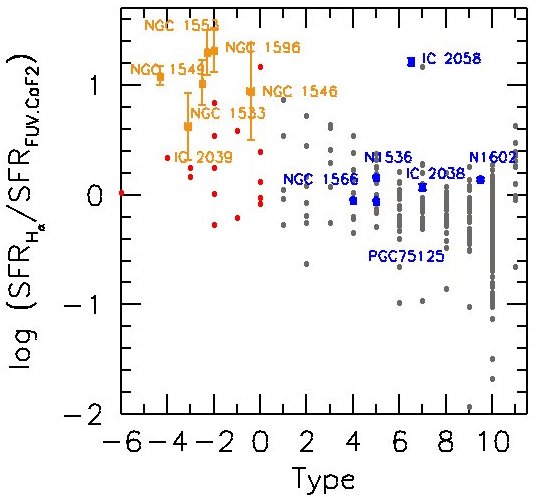}}
\caption{{\it (Left panel):} Comparison between SFR derived from \Ha and FUV.CaF2 
luminosities.  The Dorado sample is plotted with orange (ETG) and blue (LTG) 
squares. For comparison, we plot  \citet[][their Table 2]{Lee2009} sample with 
red (ETG) and gray (LTG) dots. The magenta dot shows NGC1533 as derived
using SFR from \Ha\ by \citet{Kaisina2012} and from FUV.CaF2 measured in this work. 
All measures account for Galactic extinction, but internal dust attenuation 
has not been modelled and accounted for.  
The black solid and dashed  lines show the \citet{Lee2009} relation
and the one-to-one correspondence, respectively.  The solid green and orange-dashed 
lines are the regression fit of the entire Dorado sample (equation~\ref{eq3}) and
of the ETGs members (equation~\ref{eq2}), 
respectively.
{\it (Right panel):} Ratio of SFR from \Ha and FUV.CaF2 versus galaxy 
morphological type. 
\label{SFR-1} } 
\end{figure*}

\begin{figure*}
	{\includegraphics[width=9.cm]{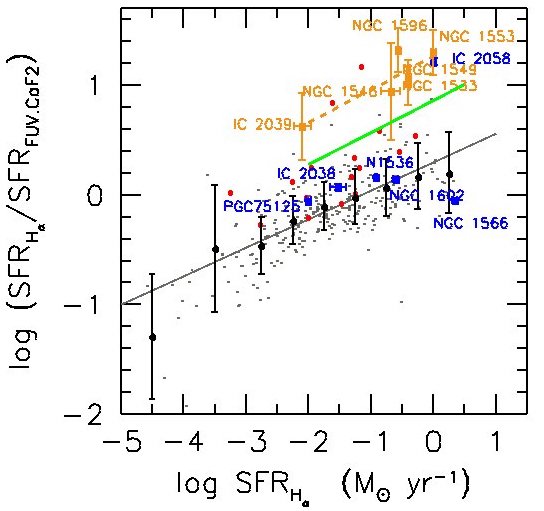}}
	{\includegraphics[width=9.cm]{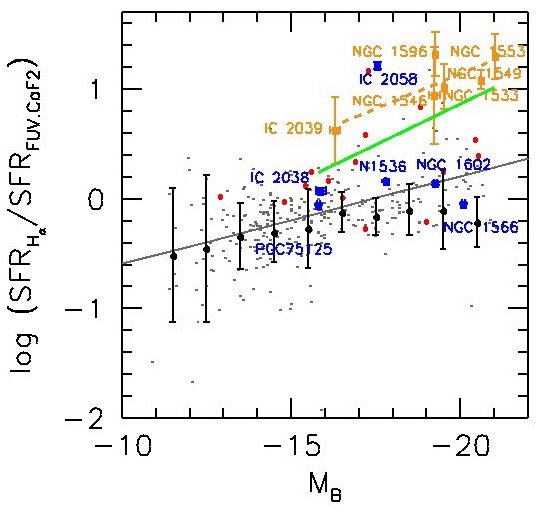}}
	\caption{Ratio between the SFR from \Ha\ and FUV.CaF2 luminosities 
  in logarithmic scale versus the SFR from \Ha\ 
(left panel) and the absolute B-band magnitude (right panel). 
Symbols used are the same as in Figure~\ref{SFR-1}. Black dots are  
average values of the sample of \citet[][their Table 2]{Lee2009}.
Solid grey lines are the linear least squares fits by \citet{Lee2009}. 
The solid green and orange dashed dashed lines are the regression fit of the 
entire Dorado sample (equation~\ref{eq4} in left panel and equation~\ref{eq6} 
in the right panel) and for the ETGs (equation~\ref{eq5} in left panel 
and equation~\ref{eq7} in the right panel), respectively.
\label{SFR-2}}
\end{figure*}

\bigskip
\noindent

We may summarize, the discussion about FUV.CaF2
extension  with respect to optical and \HI\ observations
as follows:

\begin{itemize}

\item{The radius of the FUV.CaF2 emission in Dorado ETGs is several times 
smaller than the optical one as measured either by {\tt VST} or CGS. This
is also the case of NGC 1533 (Ram2021).}

\item{Since the central FUV.CaF2 emission is always consistent with a disk structure,
the dissipative processes triggered the recent SF and exhausted the
\HI\ in NGC 1549 and NGC 1553 not revealed by the 
\citet[][see their Figure 14]{Elagali2019} mosaic of the Dorado group.}

\item{In the periphery of SGC~0414-5559, i.e NGC 1546 and in the pair 
IC 2058/PGC 75125, the FUV.CaF2 emission is coupled with \HI\ emission
\citep{Kilborn2005,Elagali2019} and on-going SF \citep[see also][]{Rampazzo2021}. }

\item{In LTGs the FUV.CaF2 and optical {\tt VST} radii are comparable
like IC 2058, NGC 1536 and NGC 1566. Larger FUV.CaF2 extension,
XUV disks, are detected in IC 2038 (Ram2021). 
In NGC 1566 the \HI\ gas emission extends well beyond the   
the FUV.CaF2 radius measured in the present study (Table~\ref{UVIT-opt}).
The fact that in NGC 1566 both optical and the FUV.CaF2 are of similar size may
indicate that the SF drops well before the \HI\ extension.} 

\item{The distortion revealed by {\tt VST} images in the outskirts
of NGC 1602 superposes to the \HI\ pipeline emission that connects it to
NGC 1596 \citep{Bureau2006,Chung2006} and may feed its recent SF 
detected by {\tt UVIT}.}

\end{itemize}

\section{SFR from FUV.CaF2 flux versus \Ha\ flux}
\label{FUV-SFR}

In this section we compare the SFR of Dorado members as derived from
\Ha   (Ram2020) with  that obtained from the
FUV.CaF2 luminosities (Section \ref{SFR_FUV}). In the
comparison we include also members of the NGC 1533 Dorado substructure 
previsouly analyzed by Ram2021, thus covering most of the backbone of
the Dorado group. We also compare the SFRs of Dorado members with that of
galaxies in the Local Volume (LV) \citep{Lee2009,Karachentsev2013}

Regarding the comparison  between the SFR obtained from different indicators,
we recall some critical points discussed by \citet{Bianchi2011}. First of all,
UV  probes ages up to hundreds of Myr, while \Ha\ 
traces ionizing photon by O and early B stars, i.e. significantly younger ages 
(Section \ref{SFR_FUV} and references therein). Therefore, the  
\citet{Kennicutt2009} relations are a good approximations for constant average 
SFR  ($t>10$ Gyr). Second, UV fluxes are affected by  reddening more severely than
optical fluxes, and, furthermore,  dust properties and extinction 
vary with the local stellar environment. Accurate modeling of galaxy properties 
from UV data must also account for internal extinction, in particular for 
LTGs \citep[][Table~1]{Bianchi2017}.\\

In Figure~\ref{SFR-1} (left panel) we compare the SFR obtained 
from \Ha\ (Ram2020) and {FUV.CaF2} fluxes (Table~\ref{UVIT-results}). 
Both fluxes are corrected for the foreground extinction
as described in Section~\ref{SFR_FUV}.
No correction for  internal extinction has been applied.
Figure ~\ref{SFR-1} shows that all  Doardo LTGs (T$>0$, blue squares) 
are essentially distributed along the 1:1 line (black-dashed line in the left panel), 
except for the edge-on IC 2058, where neglecting internal extinction 
correction can significantly alter the result.

For the Dorado  ETGs ($-5\leq T \leq 0$ orange squares),  instead,  
SFR$_{FUV.CaF2}$ is at least is 1 dex  lower than SFR$_{H\alpha}$.   
Such marked difference is also seen in the right panel of Figure~\ref{SFR-1} 
where  the ratio [SFR$_{H\alpha}$/SFR$_{FUV.CaF2}$]  is plotted as a function 
of the morphological type.

Given that the Dorado's   ETGs define a well separated locus from the LTGs, and a clear trend (roughtly paralell to that of the LTGs, but more than 1dex offset), we fit the relation between their SFR values and find: 

\begin{equation}
		\label{eq2} 
	\log \left(SFR_{FUV.CaF2}\right)= 0.70 \times \log \left(SFR_{H\alpha}\right) -1.26 
\end{equation}	

with a correlation index= 0.97. 
The regression fit is shown as a dashed orange line in Figure~\ref{SFR-1}. 

For comparison, we also plot in Figure~\ref{SFR-1} the sample of about 300 local 
volume ($<$11~Mpc) galaxies by \citet{Lee2009}, who used GALEX UV data 
and ground-based \Ha\ measurements to derive SFR; these authors account for the
foreground but not for internal extinction, as for our data-set.  
Their sample included LTGs (the majority, gray dots in our figure) 
and a few ETGs (red dots). \citet{Lee2009} derived a relation from their whole 
sample, which we report the left panel of in Figure~\ref{SFR-1} as a solid black line. 
A few facts appear evident from Figure~\ref{SFR-1}.   \citet{Lee2009} relation 
is driven by their LTGs (more numerouns in their sample), and it does not strongly 
differ from the distribution of the Dorado's LTGs; these seem more closely 
aligned along the 1:1 line (again excluding the edge-on case, that would require 
a large extinction correction); we do not derive a formal regression fit for 
the Dorado LTGs due to the scarcity of the sample. 
The ETGs in \citet{Lee2009}  sample (red dots in Figure~\ref{SFR-1}) span a 
broad locus, from their relation derived for the whole sample  all the way 
to the locus of the Dorado ETGs.  \citet{Karachentsev2013}  had also studied the 
SFR of the LV galaxies using \Ha\ and  FUV measures, and noted that ETGs show a 
large dispersion in SFR.

The SFR$_{FUV.CaF2}$ versus  SFR$_{H\alpha}$ trend for the entire Dorado sample, 
shown in Figure~\ref{SFR-1}  by a green continuous line, is represented by the relation: 

\begin{equation}
	\label{eq3}
	\log \left(SFR_{FUV.CaF2}\right)= 0.77 \times \log\left(SFR_{H\alpha}\right) -0.43  
\end{equation}

with correlation index= 0.84. The comparison of this relation with the general 
relation by \citet{Lee2009}  and with the relation for Dorado's ETGs simply 
reflects the differing composition of the samples, i.e. the relative content 
of 'classical' LTGs {\it vs} ETGs. The relations are approximately 
parallel, as expected because  they  trace a gradient of SFR intensity 
across the sample.  

This  is further evidenced in Figure~\ref{SFR-2} 
where the trend of the log(SFR$_{H\alpha}$/SFR$_{FUV}$) 
ratio is shown as a function of the log~SFR(H$\alpha$)
(left panel) and the absolute B-band magnitude, M$_B$ (right panel)
in Table~\ref{UVIT-sources}. We fit our values obtaining relations,
for the entire Dorado sample (equation~\ref{eq4}, \ref{eq6}) and for ETGs
(equation~\ref{eq5}, \ref{eq7}). The relations plotted in the left panel 
of Figure~\ref{SFR-2} are:

\begin{equation}
	\log \left(SFR_{FUV.CaF2}/SFR_{H\alpha}\right)= 0.24 \times \log\left(SFR_{H\alpha}\right)  + 0.46  
	\label{eq4} 
\end{equation}	

\begin{equation}
	\log \left(SFR_{FUV.CaF2}/SFR_{H\alpha}\right)= 0.30 \times \log \left(SFR_{H\alpha}\right) + 1.26 
	\label{eq5} 
\end{equation}	

\noindent
while in the right panel are:

\begin{equation}	
	\log \left(SFR_{FUV.CaF2}/SFR_{H\alpha}\right)= -0.15 \times M_B  - 2.11  
	\label{eq6} 
\end{equation}

\begin{equation}
	\log \left(SFR_{FUV.CaF2}/SFR_{H\alpha}\right)= -0.13 \times M_B - 1.40 
	\label{eq7} 
\end{equation}

If we exclude IC 2058, Dorado LTGs roughly align along the  
SFR$_{H\alpha}$ vs. SFR$_{FUV.CaF2}$ one-to-one correspondence
line as shown in Figure~\ref{SFR-1}. 
We conclude that  the relation between the two SFR indicators  for LTGs 
(especially at log (SFR) $\gtrsim$ -2.5)  is essentially consistent 
across different LV environments, while ETGs can depart by over one order of 
magnitude. \\

We  believe it is unlikely that internal extinction in LTGs could account 
for the large discrepancy between SFR$_{FUV}$ vs SFR$_{H\alpha}$.  
\citet{Mazzei2014a,Mazzei2014b} have performed full modeling with  
smoothed particle hydrodynamic 
simulations with chemo-photometric implementation (SPH-CPI) of 
two galaxies, one in our sample (NGC 1533) and one with similar 
characteristics (NGC1543, \citet[see][]{Marino2011a}, Ram2020). 
The simulations included self-gravity of gas, stars and dark
matter, radiative cooling, hydro-dynamical pressure, shock heating,
viscosity, star formation and feedback from evolved stars and type II
supernovae and chemical enrichment, and matched all the observed properies 
of the galaxies.  Their results showed that internal reddening could 
account for a factor of 2 correction to the SFR, at most.

We argue that the LTGs versus ETGs behaviour shown
by Figure~\ref{SFR-1} and \ref{SFR-2} is not surprising, 
given the different modality of  SF which is residual 
in ETGs at odds with LTGs.
SF episodes in ETGs cannot support a constant SF for  
10$^8$ yr or longer (1 Gyr), as required for  SFR$_{UV}$  
\citet{Kennicutt1998} calibration. 
Moreover, as \citet[][and references therein]{Kennicutt2012} 
reported, the systematic dependence of the \Ha/UV ratio in 
Figure~\ref{SFR-1} and \ref{SFR-2} may be 
produced  by temporal variations in SFRs. Such variations of SFR are
tipical of ETGs as shown by SPH-CPI simulations 
\citep[see][and references therein]{Mazzei2019}.
Since the global Dorado \Ha\ vs FUV SFR relation is driven by the 
ETGs, i.e. by the Dorado evolutionary status, we argue
that FUV vs. \Ha\ SFR relation  may be caused  by
the mix of different galaxy types.

\section{Summary and conclusions}
\label{Conclusions}

We observed with the far-ultraviolet channel (FUV.CaF2) (1300-1800 \AA) of  
{\tt ASTROSAT-UVIT} 7 fields mapping Dorado, a nearby and still clumpy group, 
extending for about 10 square degrees in the Southern Hemisphere.
We present the study of its core region,
identified by the compact group SCG 0414-5559 \citep{Iovino2002}, 
the NGC 1566/NGC 1581 and the NGC 1596/NGC 1602 substructures,
these latter observed for the first time in FUV. 
We  included in the analysis also the Dorado South-West NGC 1533 substructure
previously investigated with {\tt UVIT} and the same filter in Ram2021.\\

All galaxies observed are detected in FUV.CaF2.
We revealed an inner ring  in NGC 1549, not
detected in optical bands, and a ring in NGC 1546 which adds 
to FUV rings found in NGC1533 and NGC1543 by previous 
UV studies \citep[see][and Ram2021]{Marino2011a}.
Simulations show these resonance rings 
originate by mergers,  fly-by encounters and, in general,
galaxy-galaxy interactions \citep{Mazzei2014a,
Mazzei2014b,Eliche2018,Mazzei2018, Mazzei2019,Mazzei2022}. Structural
asymmetries in the FUV.CaF2 image of NGC 1566, PGC 75125 
and [CMI2001]4136-01 are additional  signatures of interactions.
In both IC 2058 and NGC 1596 the smoothed FUV.CaF2 emission enhances 
the presence of extra-planar features not revealed in \Ha\
(Ram2020).

We  analyzed the galaxy morphology, the surface brightness profiles 
and obtained integrated magnitudes. 
We fit the surface brightness profiles adopting a single S\'ersic law.  
Excluding NGC 1602, strongly irregular in the FUV.CaF2 band, 
the range  of the S\'ersic index, $0.76\pm0.02 \leq n \leq 2.86\pm0.28$, indicates
the presence of a disk in all the  galaxies \citep[see][]{Rampazzo2017},
and suggests that dissipation mechanisms have operated in shaping the
FUV.CaF2 structure, even in the  gas poor ETGs like NGC 1553 and NGC 1549.

We used deep, $g$ and $r$ wide field  {\tt VST} images to investigate the
presence of XUV disk by comparison with the FUV.CaF2 extension. 
We found that in ETGs the FUV.CaF2 emission is several times less extended than
the optical one. At odds, LTGs have similar FUV.CaF2 and optical dimensions. 
IC 2038 is consistent with having a XUV disk (Ram2021). 

The residual SF in ETGs of Dorado is proven by 
their FUV emission, shaped in a disk, and detection of  \HII\ regions in  
a significant fraction of them (Ram2020). 
The SFR estimates from \Ha\ and UV give consistent results for LTGs. 
The only exception is IC 2058 which is seen edge-on.
At odds, the SFR estimate from \Ha\ of Dorado ETGs is at least 
10 times higher  than the UV estimate.
We derived the relations describing the SFR$_{FUV.CaF2}$ and  
the SFR$_{H\alpha}$ trends for the entire Dorado backbone 
(equations~\ref{eq3}, \ref{eq4} and \ref{eq6}) and for 
ETGs members separately (equations~\ref{eq2}, \ref{eq5} and \ref{eq7}). 

ETGs in Dorado are numerous and drive the SFR relations we derived. 
We suggest that Dorado SFR$_{H\alpha}$ versus
SFR$_{FUV.CaF2}$ relation marks  the group evolutionary status. 
On the other hand, SF episodes in ETGs cannot support a underlying, constant SFR  
as required by general calibrations \citep{Kennicutt1998}, as 
shown by SPH-CPI simulations \citep[][and reference therein]{Mazzei2022}. 

Our  analysis in the short wavelength range (FUV.CaF2) of Dorado galaxy 
members confirms the active evolutionary status of this group 
indicated   by optical \citep{Malin1983}, radio \citep{Kilborn2005,Elagali2019}, 
and \Ha\ observations (Ram2020).

\begin{acknowledgements}
The UVIT project is collaboration between the following institutes from India: 
Indian Institute of Astrophysics (IIA),
Bengaluru, Inter University Centre for Astronomy and Astrophysics (IUCAA), 
Pune, and National Centre for
Radioastrophysics (NCRA) (TIFR), Pune, and the Canadian Space Agency (CSA). 
The detector systems are provided by
the Canadian Space Agency. The mirrors are provided by LEOS, ISRO, Bengaluru and 
the filter-wheels drives are
provided by IISU, ISRO, Trivandrum. Many departments from ISAC, ISRO, 
Bengaluru have provided direct support in
design and implementation of the various sub-systems. 
Data from Extragalactic Database (NED) 
and NASA’s Astrophysics Data System (ADS) are also used in this paper.	
We acknowledge the usage of the {\tt HyperLeda} 
database ({\tt http://leda.univ-lyon1.fr}).
\end{acknowledgements}

\bibliographystyle{aa} 
\bibliography{aa-Dorado.bib} 


\end{document}